\documentclass[aps,prl,amsmath,amssymb,nofootinbib,twocolumn]{revtex4}
\usepackage{color}
\usepackage{graphicx}% Include figure files
\usepackage{bm}% bold math
\usepackage{lipsum}
\usepackage{epsfig}
\usepackage{epstopdf}
\usepackage{colordvi}
\usepackage{float}
\usepackage{accents}

\def\ba{\begin{eqnarray}}
\def\ea{\end{eqnarray}}
\def\bc{\begin{center}}
\def\ec{\end{center}}

\begin{document}

\title{Generation of Quantum Entanglement based on Electromagnetically Induced Transparency Media}
\author{You-Lin Chuang,$^{1}$}\email{yloptics@cts.nthu.edu.tw}
\author{Ray-Kuang Lee,$^{1,2,3,4,}$}
\author{Ite A. Yu$^{3,4,}$}

\affiliation{
$^1$Physics Division, National Center for Theoretical Sciences, Hsinchu 30013, Taiwan\\
$^2$Institute of Photonics Technologies, National Tsing Hua University, Hsinchu 30013, Taiwan\\
$^3$Department of Physics, National Tsing Hua University, Hsinchu 30013, Taiwan\\
$^4$Center for Quantum Technology, Hsinchu 30013, Taiwan}

\begin{abstract}
Quantum entanglement is an essential ingredient for the absolute security of quantum communication. Generation of continuous-variable entanglement or two-mode squeezing between light fields based on the effect of electromagnetically induced transparency (EIT) has been systematically investigated in this work. Here, we propose a new scheme to enhance the degree of entanglement between probe and coupling fields of coherent-state light by introducing a two-photon detuning in the EIT system. This proposed scheme is more efficient than the conventional one, utilizing the dephasing rate of ground-state coherence, i.e., the decoherence rate to produce entanglement or two-mode squeezing which adds far more excess fluctuation or noise to the system. In addition, maximum degree of entanglement at a given optical depth can be achieved with a wide range of the coupling Rabi frequency and the two-photon detuning, showing our scheme is robust and flexible. It is also interesting to note that while EIT is the effect in the perturbation limit, i.e. the probe field being much weaker than the coupling field and treated as a perturbation, there exists an optimum ratio of the probe to coupling intensities to achieve the maximum entanglement. Our proposed scheme can advance the continuous-variable-based quantum technology and may lead to applications in quantum communication utilizing squeezed light.
\end{abstract}


\maketitle

%%%%%%%%%%%%%%%%%%%%%%%%%%%%%%%%%%%%%%%%%%%%%%%
\newcommand{\FigOne}{
\begin{figure}[t]
\begin{center}
\includegraphics[trim = 0 50 0 -20, scale=0.5]{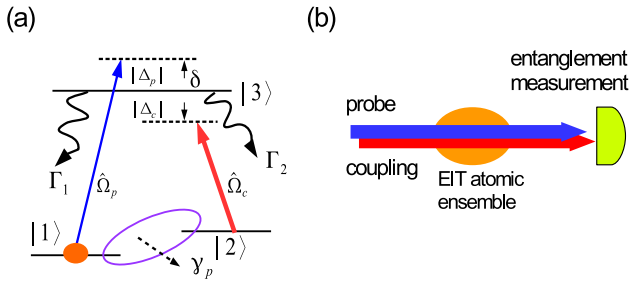} 
\end{center}
\caption{
(a) Atomic system configuration. 
(b) Probe and coupling are interacting with EIT atomic ensemble, and the entanglement measurement is performed at output.
\label{fig1}}
\end{figure}}
%%%%%%%%%%%%%%%%%%%%%%%%%%%%%%%%%%%%%%%%%%%%%%%
\newcommand{\FigTwo}{
\begin{figure}[h]
\includegraphics[scale=0.285]{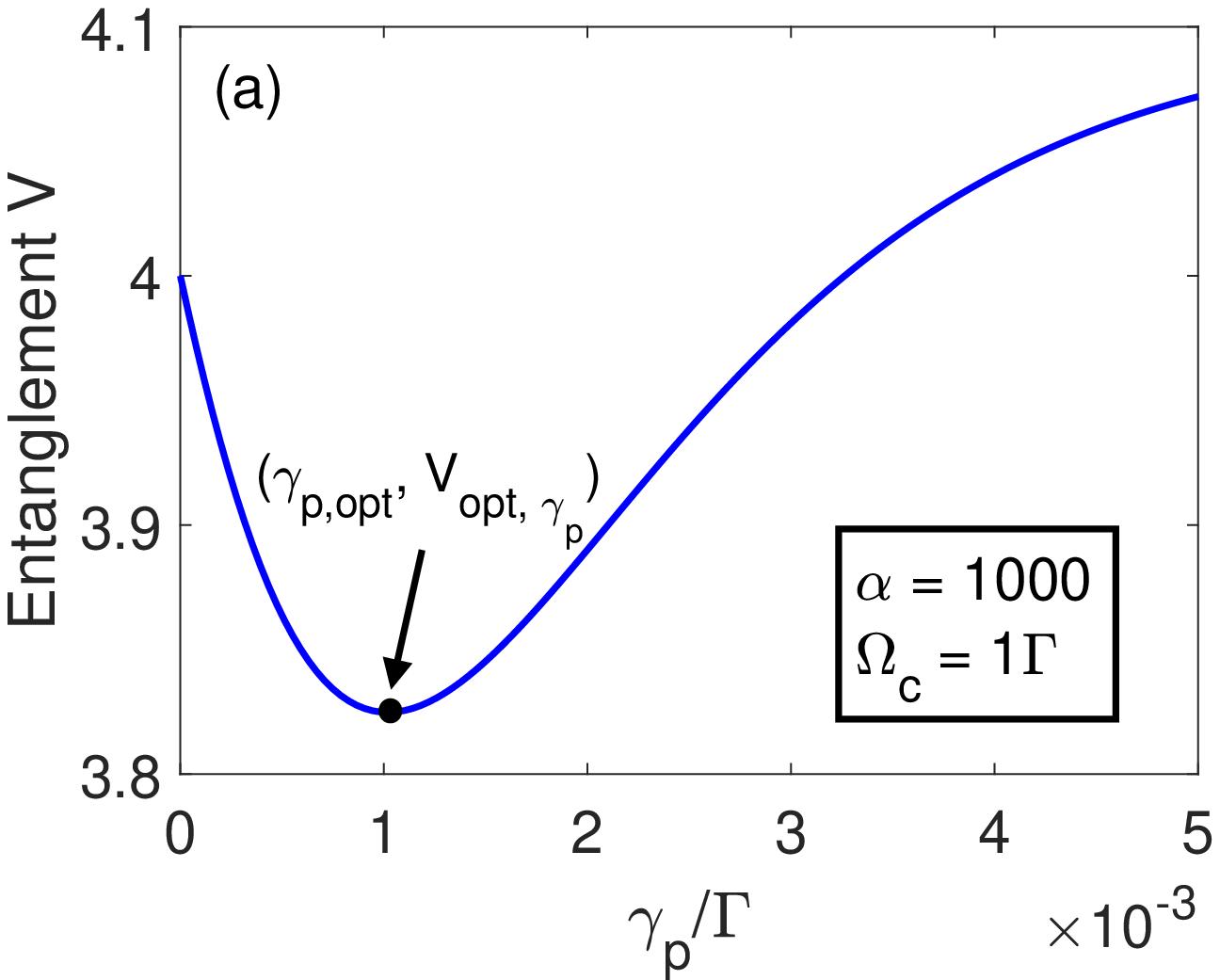} 
\includegraphics[scale=0.285]{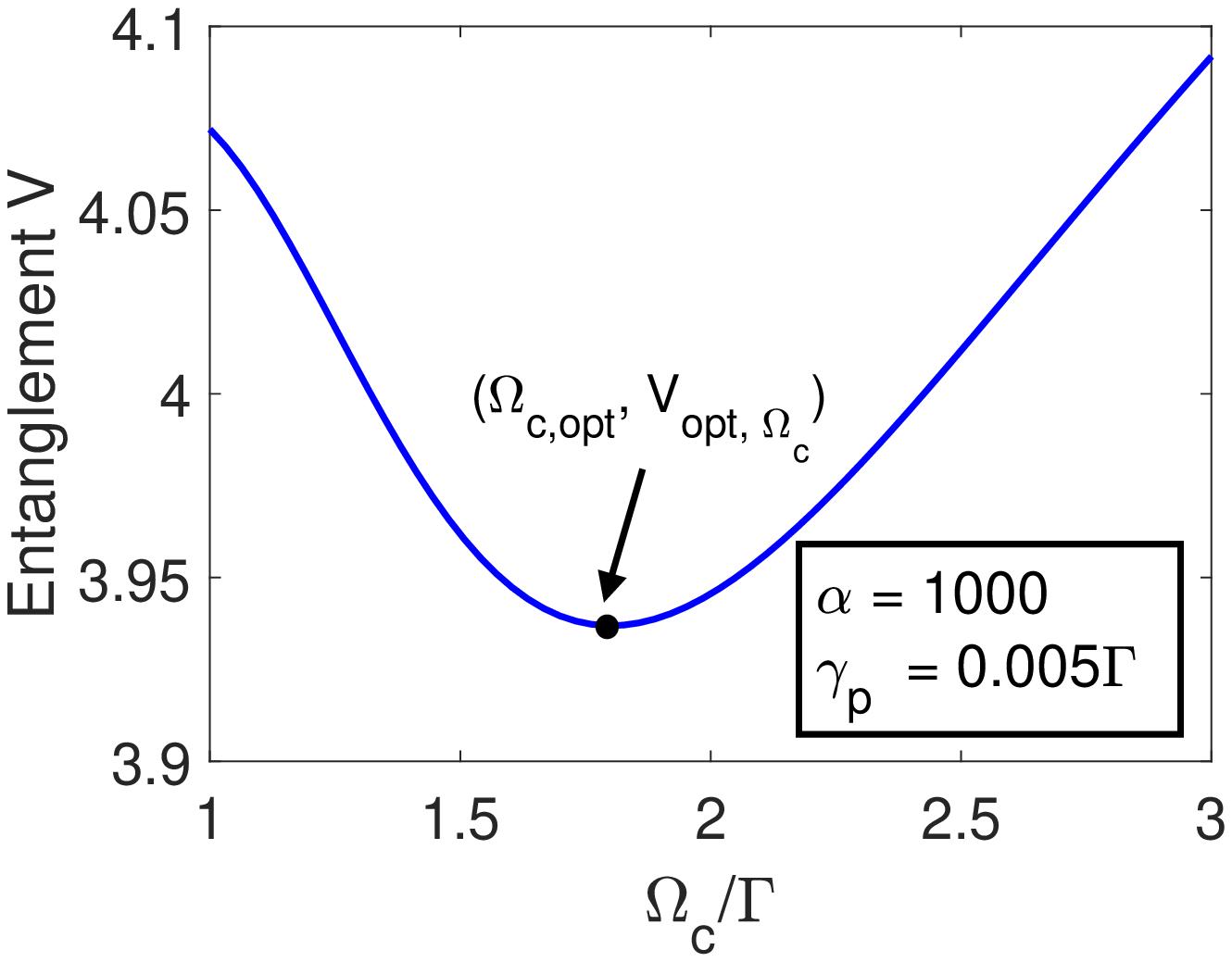} 
\caption{(a) Entanglement versus decoherence rate under the parameters given by $ \alpha = 1000 $ and $ \Omega_c = 1\Gamma$.
(b) Entanglement versus input coupling Rabi frequency with parameters given by $ \alpha = 1000 $ and $ \gamma_p  = 0.005\Gamma $.
The input Rabi frequency of probe field is used by setting $ \Omega_p = 0.1\Omega_c $, and $ \delta = 0 $ for the two figures.
\label{fig2}}
\end{figure}}
%%%%%%%%%%%%%%%%%%%%%%%%%%%%%%%%%%%%%%%%%%%%%%%
\newcommand{\FigThree}{
\begin{figure}[b]
\includegraphics[scale=0.285]{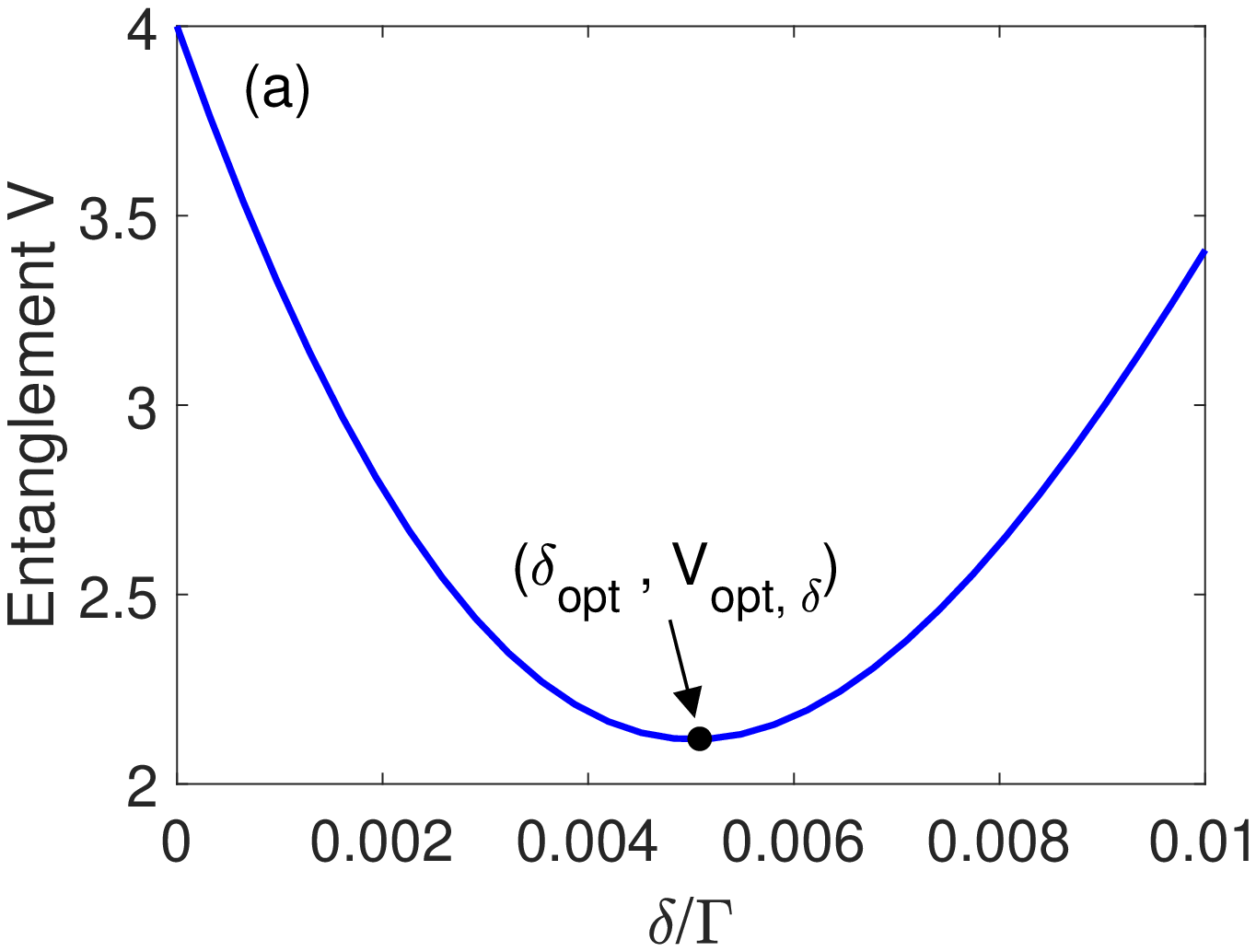} 
\includegraphics[scale=0.285]{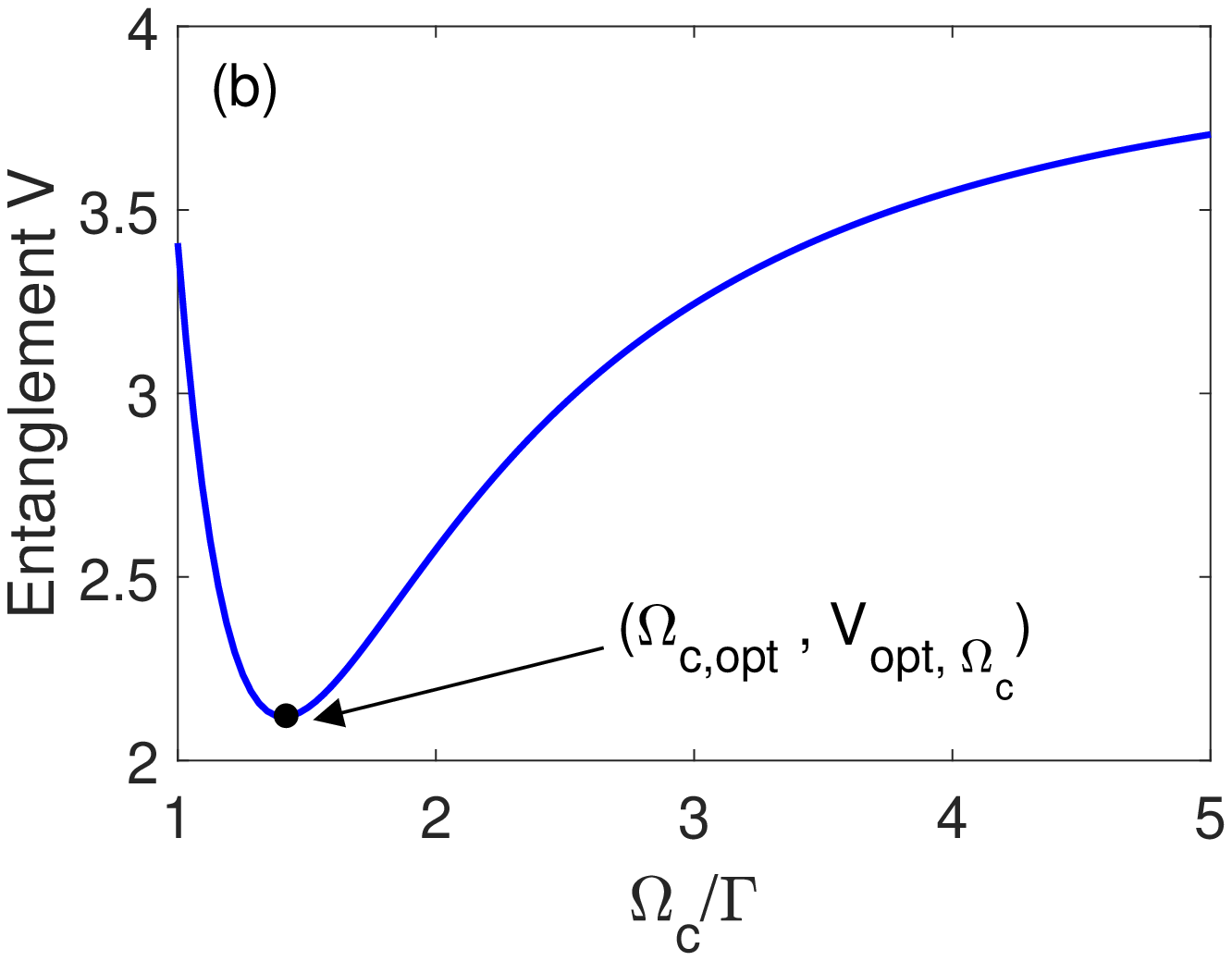} 
\caption{(a) Entanglement versus two-photon detuning under the parameters given by $ \alpha = 1000 $ and $ \Omega_c = 1\Gamma$.
(b) Entanglement versus input coupling Rabi frequency with parameters given by $ \alpha = 1000 $ and $ \delta=0.01\Gamma$.
The input Rabi frequency of probe field is used by setting $ \Omega_p = 0.1\Omega_c $, and $ \gamma_p = 0 $ for the two figures.
\label{fig3}}
\end{figure}}
%%%%%%%%%%%%%%%%%%%%%%%%%%%%%%%%%%%%%%%%%%%%%%%
\newcommand{\FigFour}{
\begin{figure}[t]
\includegraphics[scale=0.285]{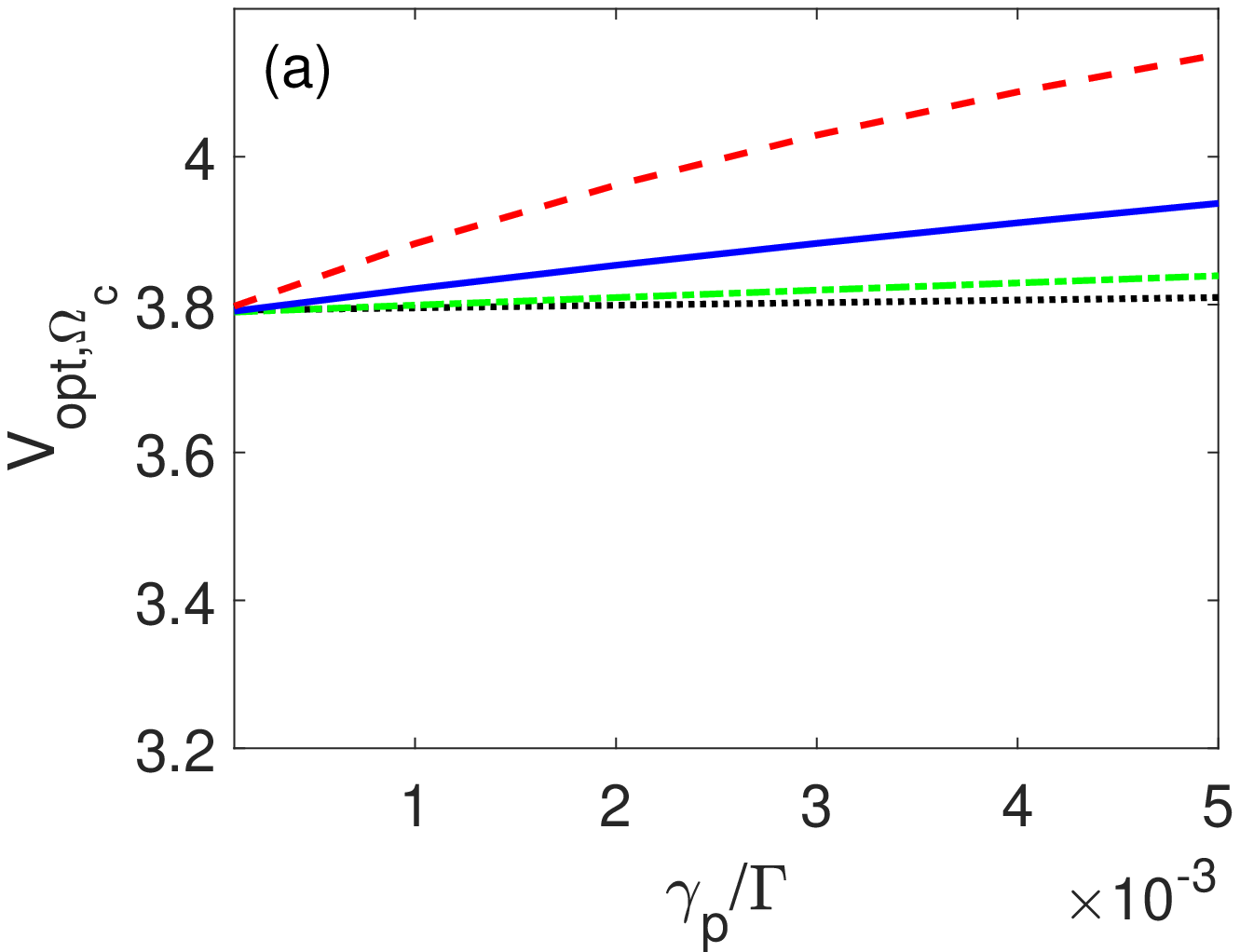} 
\includegraphics[scale=0.285]{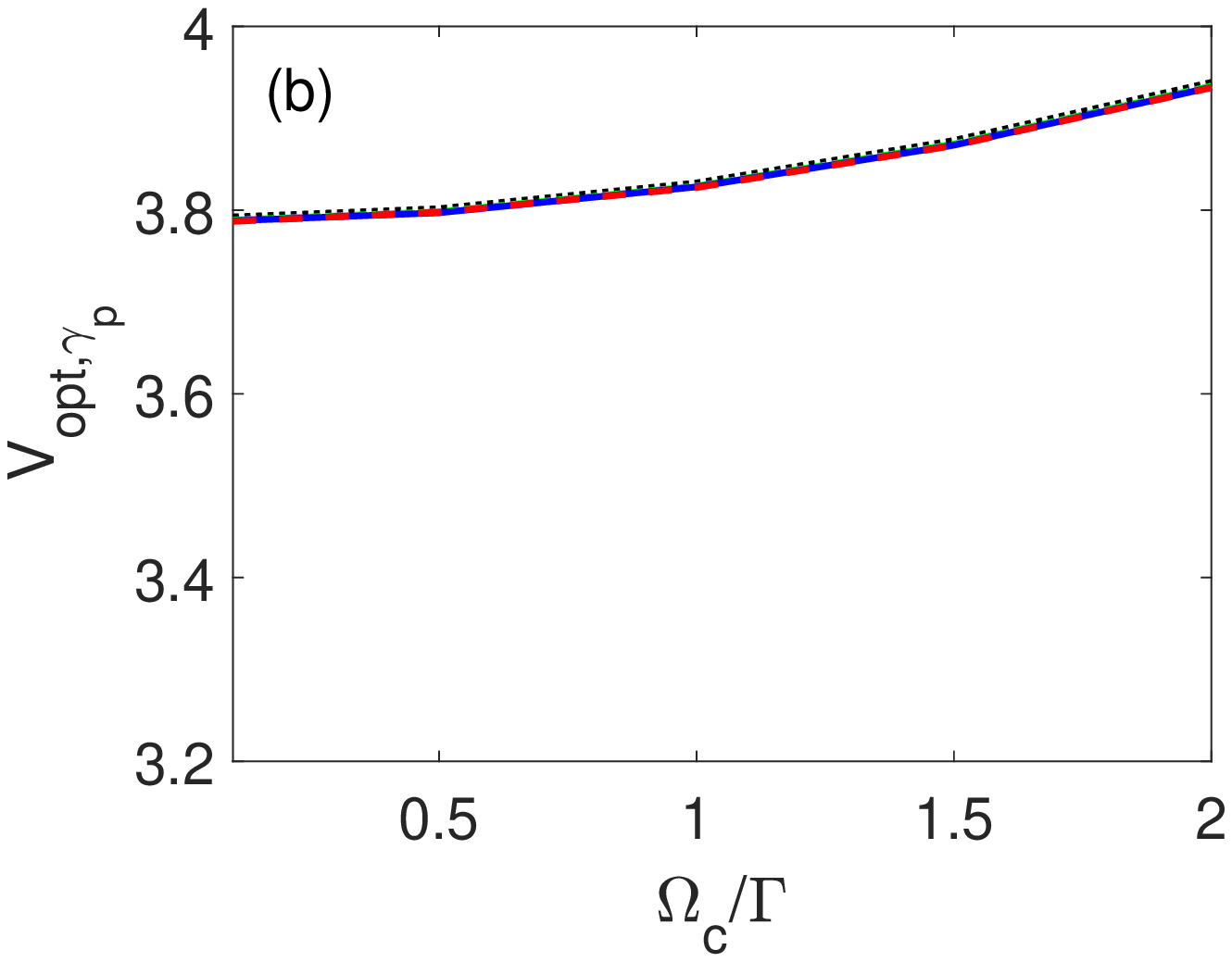} \\
\includegraphics[scale=0.285]{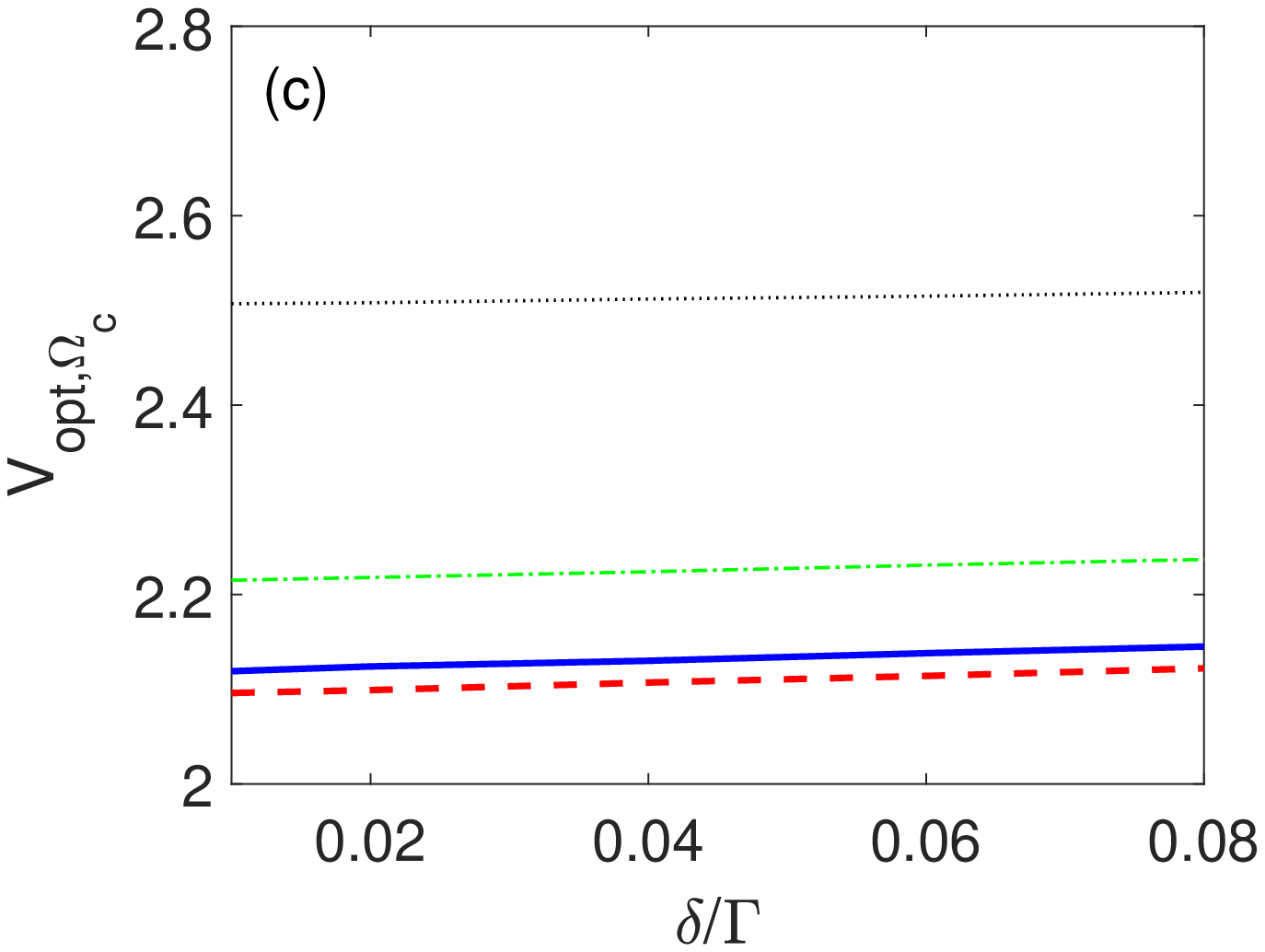} 
\includegraphics[scale=0.285]{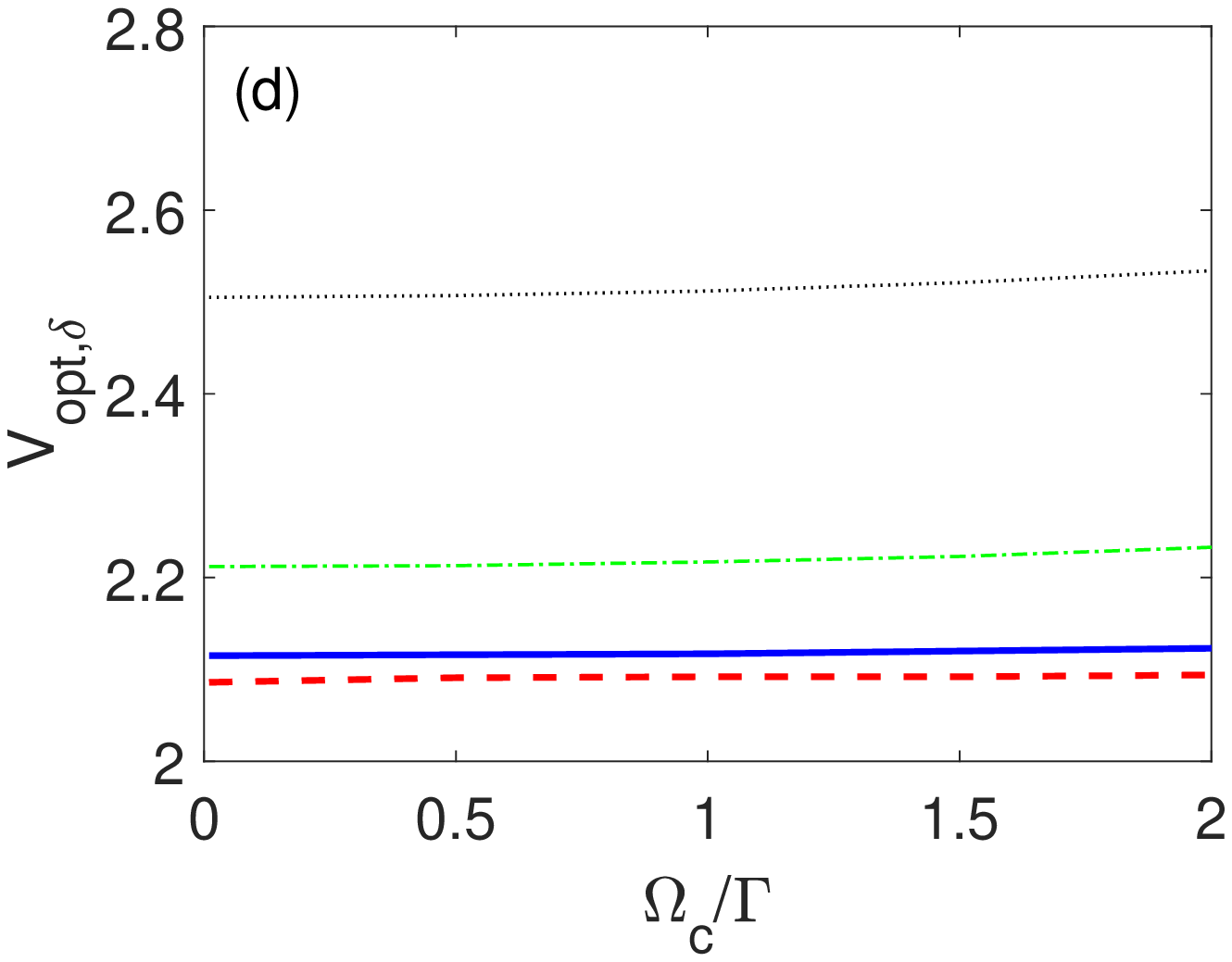} 
\caption{Optimized entanglement at different $ \alpha $'s for the two different processes: by dephasing rate of ground state coherence (a) and (b), and by two-photon detuning (c) and (d). 
Black dotted, green dashed-dotted, blue solid, and red dashed lines represnt $ \alpha $'s values of 100, 300, 1000, and 3000, respectively.
(a) The minimum value of entanglement quantity obtained by scanning all input Rabi frequencies, $ V_{\text{opt},\Omega_c} $, as a function of decoherence rate $ \gamma_p $. (b) The minimum value of entanglement quantity obtained by scanning all decoherence rate, $ V_{\text{opt},\gamma_p} $, as a function of input Rabi frequency $ \Omega_c $. 
(c) The minimum entanglement quantity obtained by scanning all input Rabi frequencies, $ V_{\text{opt},\Omega_c} $, as a function of two-photon detuning $ \delta $. (d) The minimum entanglement quantity obtained by scanning all two-photon detunings, $ V_{\text{opt},\delta} $, as a function of input Rabi frequency $ \Omega_c $.
$ \Omega_p = 0.1\Omega_c $ is used in all the figures.
\label{fig4}}
\end{figure}}

%%%%%%%%%%%%%%%%%%%%%%%%%%%%%%%%%%%%%%%%%%%%%%%%
\newcommand{\FigFive}{
\begin{figure}[t]
\begin{center}
\includegraphics[scale=0.55]{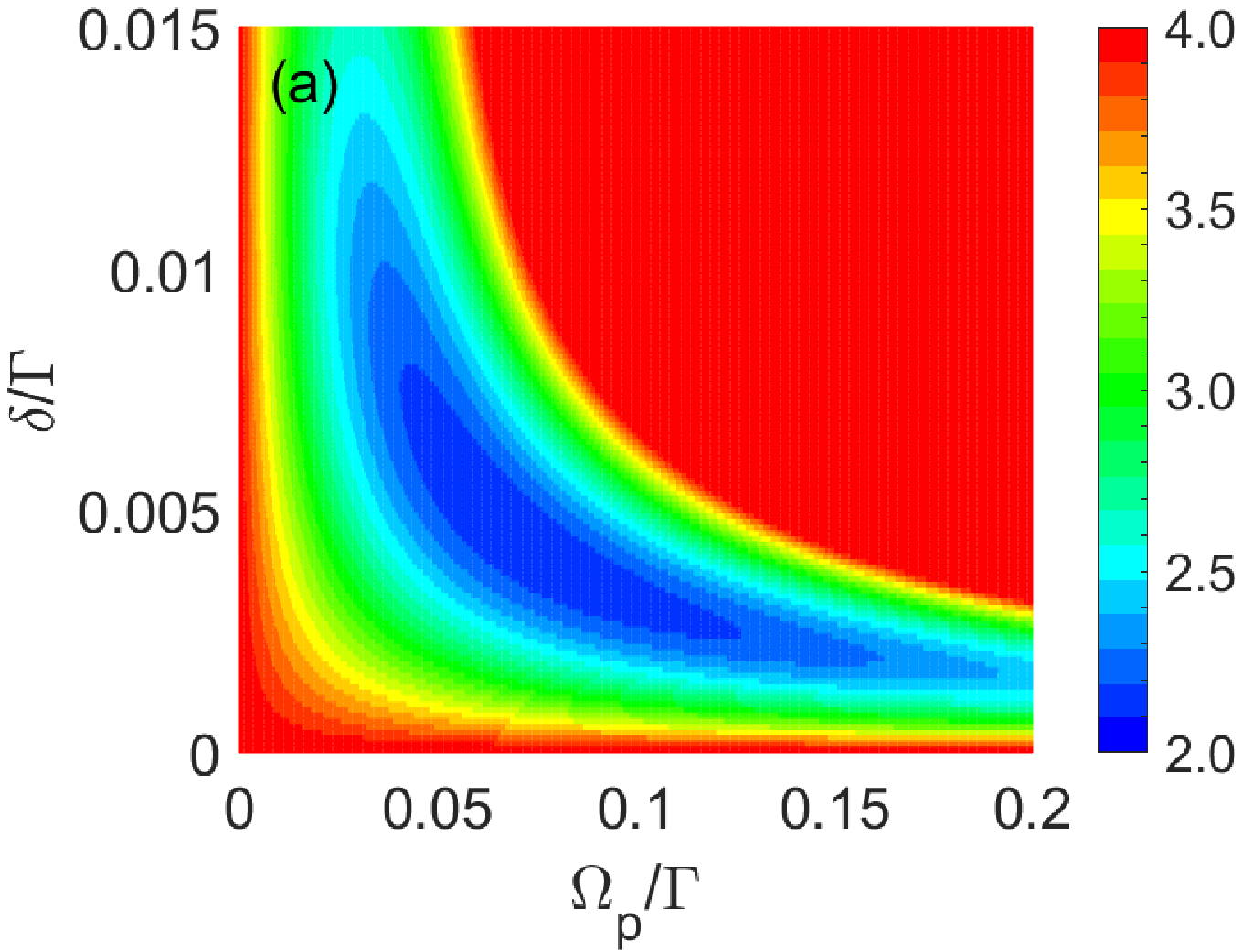} 
\includegraphics[scale=0.55]{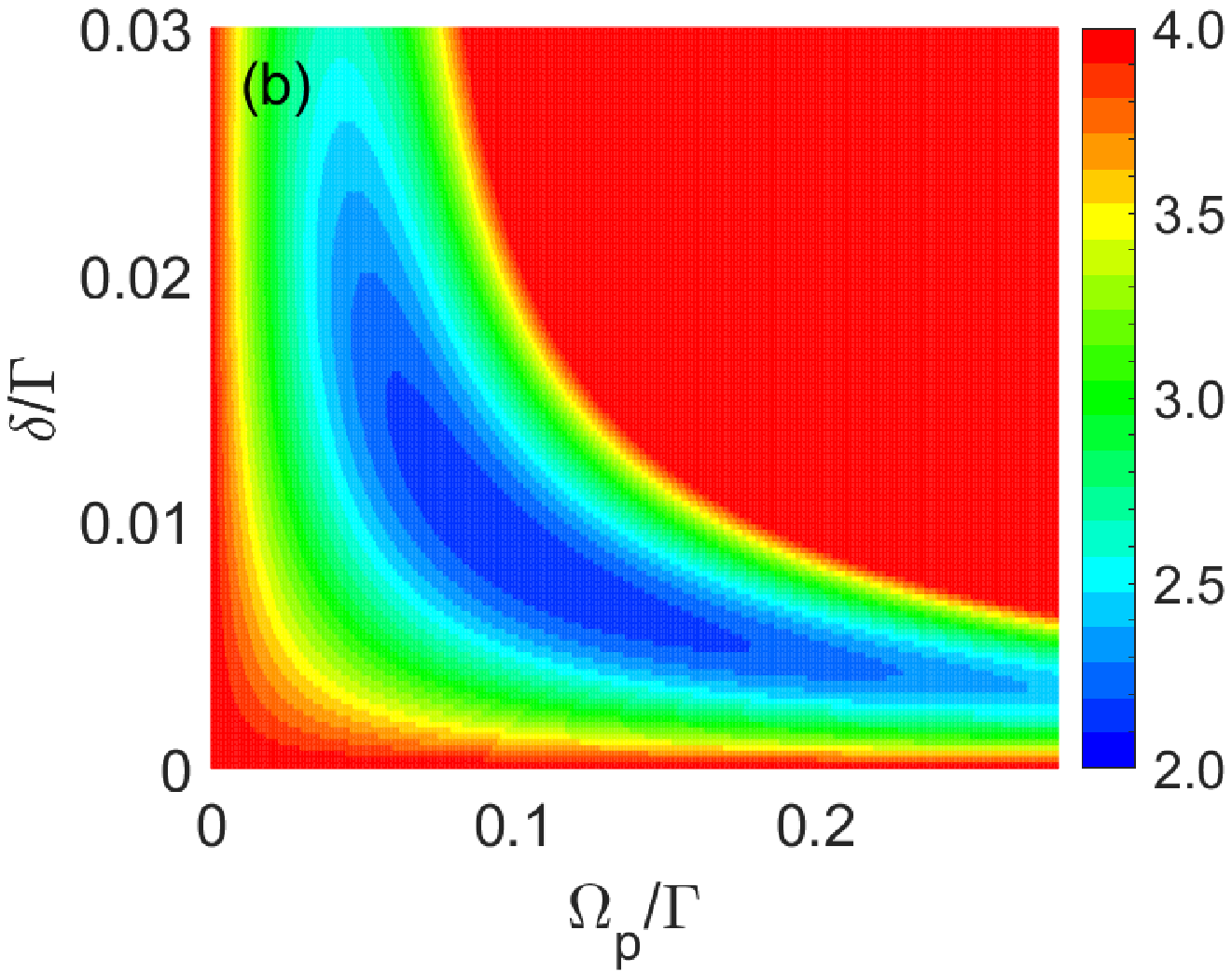} 
\includegraphics[scale=0.55]{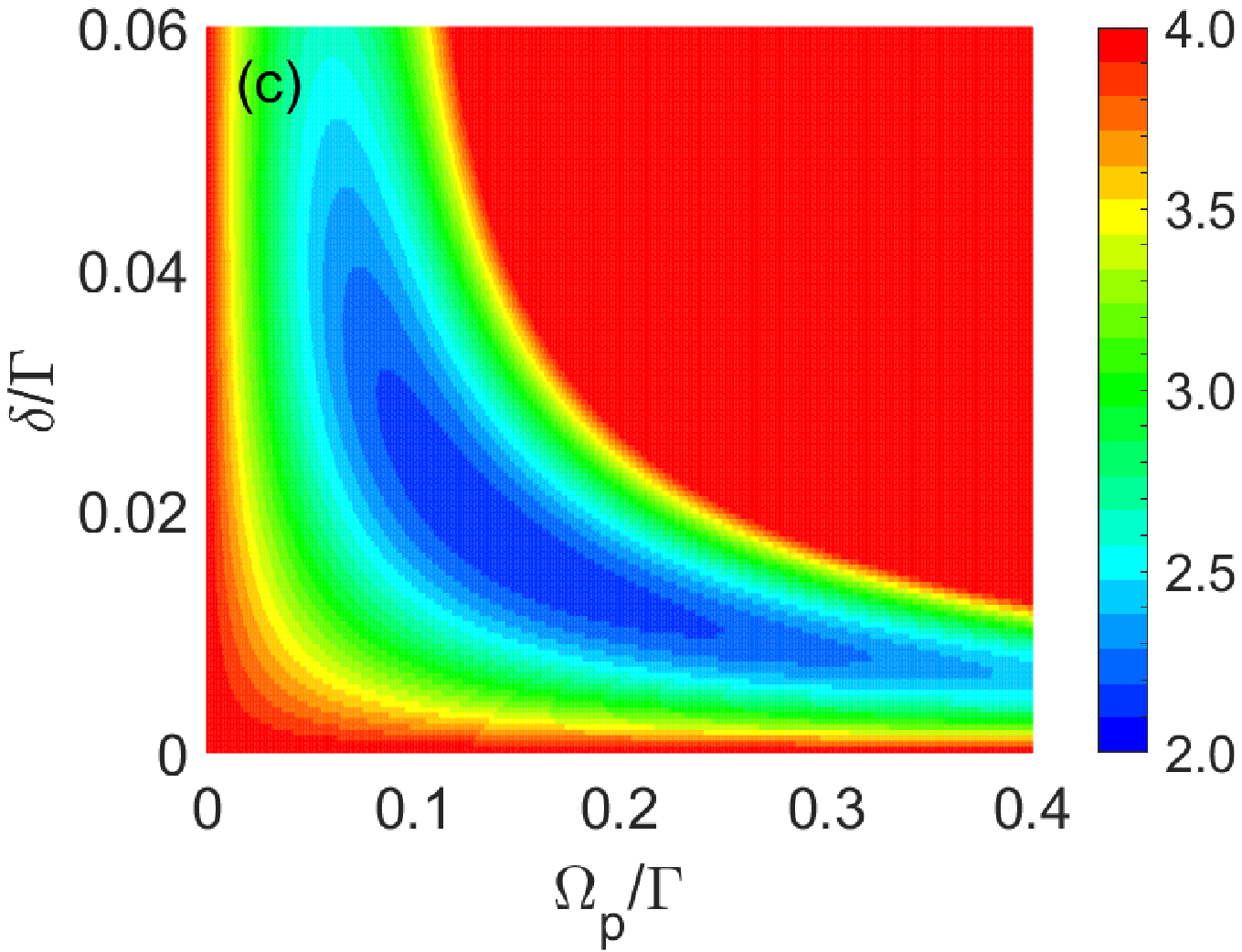} 
\end{center}
\caption{Contour plot of entanglement quantity $ V $ versus $ \Omega_p $ and $ \delta $ under different $ \Omega_c $'s: (a) $ \Omega_c = 0.85\Gamma $, (b) $ \Omega_c = 1.2\Gamma $, and (c) $ \Omega_c = 1.7\Gamma $. For the three plots, optical density is set to be $ 1,000 $. Please note that the ranges of $ \Omega_p $ and $ \delta $ are different in the three plots.  
\label{fig5}}
\end{figure}
}
%%%%%%%%%%%%%%%%%%%%%%%%%%%%%%%%%%%%%%%%%%%%%%%%
\newcommand{\FigSix}{
\begin{figure}[b]
\begin{center}
\includegraphics[scale=0.55]{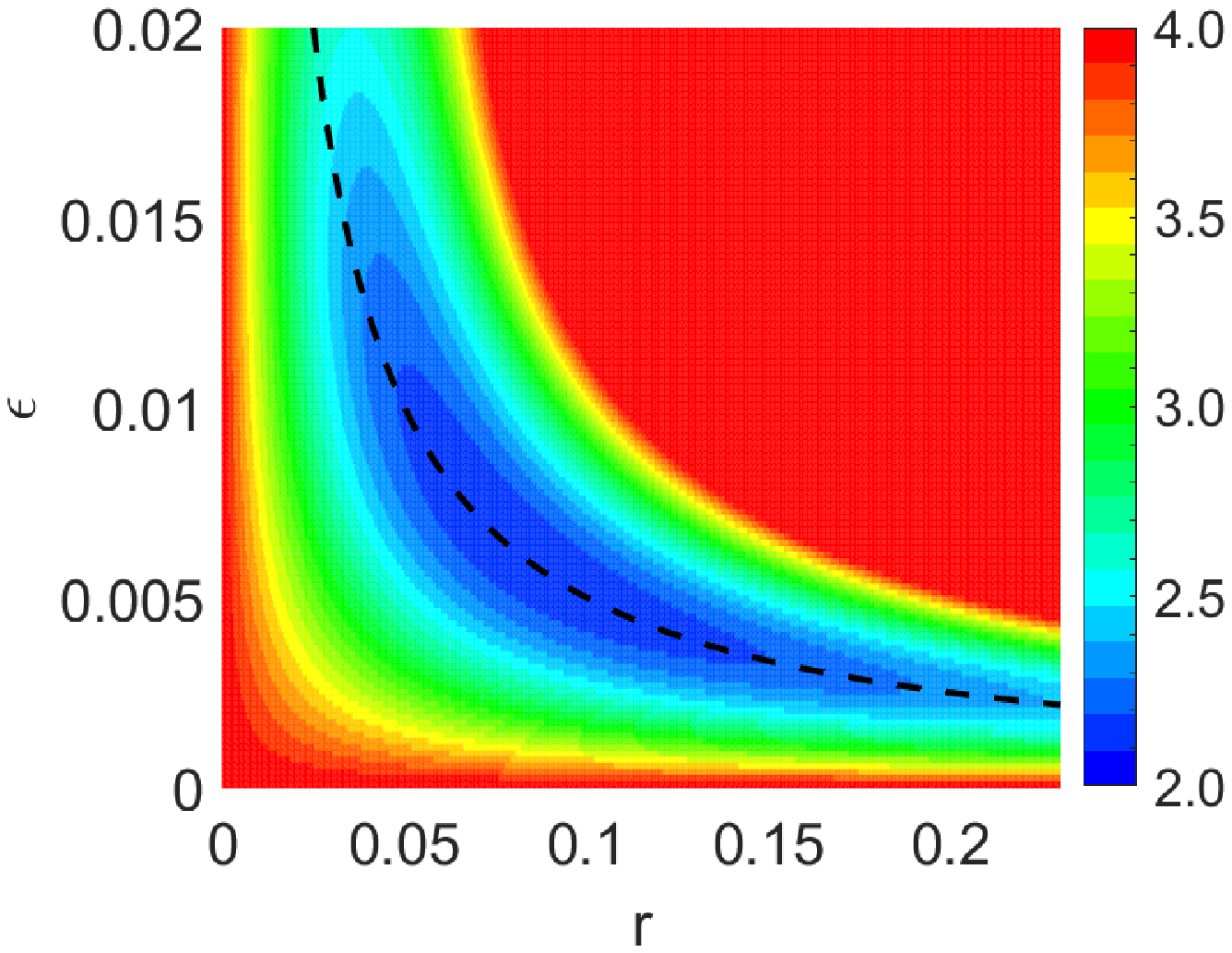} 
\end{center}
\caption{Contour plot of entanglement quantity $ V $ versus $ r $ and $ \epsilon $. The optical density is given by $ \alpha = 1,000 $. The black dashed curve is plotted with the condition of $ \mu = \alpha\epsilon r = 1/2 $.
\label{fig6}}
\end{figure}
}

\section{i. Introduction}

Continuous-variable (CV) quantum entanglement is an important resource which has been paid great attention in modern quantum optics and quantum information sciences, possessing many potential  applications in quantum teleportation \cite{qteleportation}, quantum key distribution \cite{qkeydis}, quantum communication \cite{qcommunication1, qcommunication2}, quantum information processing \cite{qinformation}, etc. Carrying quantum information onto the quadratures of optical fields, such as amplitude and phase, has higher tolerance to dissipation during light propagation processes. In addition, CV quantum entanglement can be realized 
in other degree of freedom of optical fields, for instance, the polarization state of light has also been extensively studied in CV regime by transforming the quadrature entanglement onto polarization basis \cite{PE1, PE2, PE3, ent_two_mode}, and the quadrature entanglement using quantum orbital angular momentum with spatial Laguerre-Gauss mode has been discussed in experiment \cite{CV_OAM}. 
Furthermore, CV entanglement light source is essential in quantum imaging \cite{Qimage1, Qimage2, Qimage3}, which is an extension of quantum nature to transverse spatial degree of freedom. According to these previous researches, it is believed that optical field in CV entanglement plays an ideal information carrier, which is robust in quantum information sciences. 

In order to generate entangled light, it is well known that using of optical nonlinear crystal is a typical scheme to generate light sources in CV regime. In theory, quantum correlation based on nondegenerate parametric oscillation was proposed \cite{q_correlation}. Later, the generation of CV entanglement with nondegenerate parametric amplification was first observed in experiment by Ou \textit{et al} in 1992 \cite{EPR_CV}. 
Recently, the CV quantum entanglement at a telecommunication wavelength of 1550 nm had been realized using nondegenerate optical parametric amplifer \cite{telecom_ent}. On the other hand, mixing two independent squeezed lights which are generated from optical parametric amplifiers individually provides a practical method to generate quadrature entanglement \cite{CVent_exp}. These studies above clearly indicate that there is a connection between nonlinear optical processes and CV entanglement generation, so that the integration of all optical elements on chip has been proposed in order to further approach the goal of implementation of quantum computer in future \cite{on_chip}.    

Although the generation of quantum light sources from optical parametric processes, especially using $ \chi^{(2)} $ optical susceptibility of nonlinear crystal is popular, the light-matter interaction strength is difficult to control. 
In contrast, the nonlinear optical processes based on the interactions between fields and atomic systems can produce not only large amounts of quantum correlations between intense fields, but also controllability with accessible physical parameters. Recent decades, research reported on entangled light generation by four-wave mixing (FWM) has been intensely studied in hot atomic vapors \cite{OL32, lowfrequency_qc, amplificationCV, FWM_entanglement, FWM_nonamplification, Squeezing_absorbing, qc_hotvapor}. Moreover, it has some potential applications, including the production of  multiple quantum correlated beams \cite{multiple_qc}, enhancement of the degree of entanglement \cite{cascade_enhancement}, quantum metrology \cite{metrology_PA}, etc. Meanwhile, electromagnetically induced transparency (EIT) \cite{EIT} which is a coherent process also plays an important role in atom-field interactions. Some peculiar features such as low-absorption, slow-light \cite{SL} and quantum memory \cite{LS, q_memory, Yu}  make EIT to be a promising ingredient in the development of quantum technologies. In the scenario, quantum optical pulse propagation in EIT \cite{qEIT}, quantum squeezing generation in coherent population trapping media \cite{CPTsq}, large cross-phase modulation at few-photon level \cite{XPM}, and quantum correlated light generation as well as multiple fields correlation have been actively studied \cite{superPoissionian, EITcorrelation1, EITcorrelation2, EITcorrelation3, XPM_eit, Q_ent_ultraslow, spin_coherence1, spin_coherence2, entangler, EIE}. 

CV quantum entanglement arising from atom-field interactions is a good platform to investigate the connections between quantum coherence and correlations. Despite the fact that many papers have discussed the quantum entanglement generation in EIT systems, a systematic understanding of the physics behind the entanglement generation is still lacking. In this paper, we will discuss about the following questions: how do the tunable physical parameters in EIT system, which are photon-detunings, field Rabi frequencies, and atomic optical density, influence the entanglement degree between interacting fields?  And how is the entanglement affected by the ``EIT degree", which is related to ratio between two interacting field strengths. By solving the coupled equations of atomic and field operators numerically, we are able to study these questions.

The paper has been organized in the following way.
In Sec.~II, we start from a standard analysis of typical EIT interaction Hamiltonian, and derive the equations of motion for atomic operators, as well as the propagation equations for two quantized fields.
The results from numerical calculation  are given in Sec.~III. Then, to reveal the underline physics, Sec.~IV is concerned with the analytical approach for the output entanglement. Finally, a conclusion is given in Sec. V.

\section{ii. Theoretical Model}

\FigOne

We consider a collection of atoms having the three-level $ \Lambda $-type configuration as shown in Fig.~\ref{fig1}(a). 
Two ground states $ \vert 1 \rangle $ and $ \vert 2 \rangle $ are coupled to the common excited state $ \vert 3 \rangle $ by probe and coupling fields, respectively. The Rabi frequency of probe field $ \Omega_p $ is much weaker than that of the coupling $ \Omega_c $, and  the whole system forms a standard EIT. The one-photon detuning of probe and coupling fields are defined by $ \Delta_p = \omega_p - \omega_{31} $ and $ \Delta_c = \omega_c - \omega_{32} $, where $ \omega_p $ and $ \omega_c $ denote the light frequencies of probe and coupling, and 
$ \omega_{\mu\nu} \equiv (E_\mu - E_\nu)/\hbar $ is the energy difference between any two states $ \vert \mu \rangle $ and $ \vert \nu \rangle $. The two-photon detuning is an important parameter in EIT system, defined by $ \delta = \Delta_p - \Delta_c $.
The decay rates from $ \vert 3\rangle $ to the two ground states are $ \Gamma_{1} $ and $ \Gamma_{2} $, which are assumed to be the same in this work, so that we have $ \Gamma_{1} = \Gamma_{2} = \Gamma/2 $, where $ \Gamma $ is the total decay rate of $ \vert 3\rangle $. 
In addition, the dephasing rate of ground state coherence between $ \vert 2 \rangle $ and $ \vert 1 \rangle $ is $ \gamma_p $,  i.e., the decoherence rate.

The system is arranged as shown in Fig.~\ref{fig1}(b). Probe and coupling fields are propagating along the same direction, illuminating the EIT atomic ensemble, which is cooled to several hundred micro-Kelvin. Both probe and coupling fields are coherent states at input, and the entanglement measurement is performed at output for the two fields after propagating though EIT ensemble.  

Next, we start to study the system theoretically. We write the atom-field interaction Hamiltonian $ \hat{H} $ in the rotating wave approximation
\begin{eqnarray}
\hat{H} &=& -\hbar\left( \Delta_p \hat{\sigma}_{33}(z,t) + \delta \hat{\sigma}_{22}(z,t) \right) \\\nonumber
&-& \hbar \left[ \dfrac{\hat{\Omega}_p(z,t)}{2}\hat{\sigma}_{31}(z,t) + \dfrac{\hat{\Omega}_c(z,t)}{2}\hat{\sigma}_{32}(z,t) + H.c \right] \label{H},
\end{eqnarray}  
in which $ \hat{\Omega}_p(z,t) = g_p\hat{\mathcal{E}}_p(z,t) $ and $ \hat{\Omega}_c(z,t) = g_c\hat{\mathcal{E}}_c(z,t) $, with $ g_p = \mu_{13}\sqrt{\omega_p/2\epsilon_0 V \hbar} $ and $ g_c = \mu_{23}\sqrt{\omega_c/2\epsilon_0 V \hbar} $, are the single photon Rabi frequencies of probe and coupling fields, respectively, corresponding to two dipole transitions $ \mu_{13} $ and $ \mu_{23} $. Without loss of generality, we have assumed that $ g_p = g_c \equiv g $, and $ \hat{\mathcal{E}}_p $ and $ \hat{\mathcal{E}}_c $ are dimensionless field operators, which satisfy bosonic commutation relations given by
$ \left[ \hat{\mathcal{E}}_\mu,\hat{\mathcal{E}}_\mu^{\dagger}\right] = 1  $, $ \mu \in {p,c} $.
According to the Hamiltonian in Eq.~(\ref{H}), we can write down the Heisenberg-Langevin equations for atomic operators.  
\begin{widetext}
\begin{eqnarray}
&&\dfrac{\partial}{\partial t}\hat{\sigma}_{31} = -\left( \dfrac{\Gamma}{2}+i\Delta_p\right)  \hat{\sigma}_{31}-\dfrac{i}{2}(\hat{\sigma}_{11}-\hat{\sigma}_{33})\hat{\Omega}_p^{\dagger} - \dfrac{i}{2}\hat{\Omega}_c^{\dagger}\hat{\sigma}_{21}+\hat{F}_{31}\label{S31},\\
&&\dfrac{\partial}{\partial t}\hat{\sigma}_{32} = -\left( \dfrac{\Gamma}{2}+i\Delta_c\right) \hat{\sigma}_{32}-\dfrac{i}{2}(\hat{\sigma}_{22}-\hat{\sigma}_{33})\hat{\Omega}_c^{\dagger} - \dfrac{i}{2}\hat{\Omega}_p^{\dagger}\hat{\sigma}_{12}+\hat{F}_{32}\label{S32},\\
&&\dfrac{\partial}{\partial t}\hat{\sigma}_{21} = -\left( {\gamma}_p+i\delta\right) \hat{\sigma}_{21}+\dfrac{i}{2}\hat{\Omega}_p^{\dagger}\hat{\sigma}_{23} - \dfrac{i}{2}\hat{\sigma}_{31}\hat{\Omega}_c+\hat{F}_{21}\label{S21},\\
&&\dfrac{\partial}{\partial t}\hat{\sigma}_{11} = \Gamma_1 \hat{\sigma}_{33} -\dfrac{i}{2}\hat{\sigma}_{31}\hat{\Omega}_p + \dfrac{i}{2} \hat{\Omega}_p^{\dagger}\hat{\sigma}_{13}+\hat{F}_{11}\label{S11},\\
&&\dfrac{\partial}{\partial t}\hat{\sigma}_{22} = \Gamma_2 \hat{\sigma}_{33}-\dfrac{i}{2}\hat{\sigma}_{32}\hat{\Omega}_c + \dfrac{i}{2} \hat{\Omega}_c^{\dagger}\hat{\sigma}_{23}+\hat{F}_{22}\label{S22},\\
&&\dfrac{\partial}{\partial t}\hat{\sigma}_{33} = -\Gamma \hat{\sigma}_{33}+\dfrac{i}{2}\hat{\sigma}_{31}\hat{\Omega}_p+\dfrac{i}{2}\hat{\sigma}_{32}\hat{\Omega}_c -\dfrac{i}{2} \hat{\Omega}_p^{\dagger}\hat{\sigma}_{13}-\dfrac{i}{2} \hat{\Omega}_c^{\dagger}\hat{\sigma}_{23}+\hat{F}_{33}\label{S33},\\
&&\dfrac{\partial}{\partial t}\hat{\sigma}_{12} = -\left( {\gamma}_p-i\delta\right) \hat{\sigma}_{12}-\dfrac{i}{2}\hat{\sigma}_{32}\hat{\Omega}_p + \dfrac{i}{2}\hat{\Omega}_c^{\dagger}\hat{\sigma}_{13}+\hat{F}_{12}\label{S12},\\
&&\dfrac{\partial}{\partial t}\hat{\sigma}_{23} = -\left( \dfrac{\Gamma}{2}-i\Delta_c\right) \hat{\sigma}_{23}+\dfrac{i}{2}(\hat{\sigma}_{22}-\hat{\sigma}_{33})\hat{\Omega}_c + \dfrac{i}{2}\hat{\sigma}_{21}\hat{\Omega}_p+\hat{F}_{23}\label{S23},\\
&&\dfrac{\partial}{\partial t}\hat{\sigma}_{13} = -\left( \dfrac{\Gamma}{2}-i\Delta_p\right) \hat{\sigma}_{13}+\dfrac{i}{2}(\hat{\sigma}_{11}-\hat{\sigma}_{33})\hat{\Omega}_p + \dfrac{i}{2}\hat{\sigma}_{12}\hat{\Omega}_c+\hat{F}_{13}\label{S13},
\end{eqnarray}
\end{widetext}
in which $ \hat{F}_{\mu\nu} $ is the corresponding Langevin noise operator.  $ \gamma_p $ is the dephasing rate of ground-state coherence, i.e., the decoherence rate.
The field propagations follow the Maxwell's equations given by
\begin{eqnarray}
&&\left( \dfrac{1}{c}\dfrac{\partial}{\partial t} + \dfrac{\partial}{\partial z}\right) \hat{\Omega}_p = i \left( \dfrac{\Gamma \alpha}{2L}\right)  \hat{\sigma}_{13}, \label{op}\\  
&&\left( \dfrac{1}{c}\dfrac{\partial}{\partial t} + \dfrac{\partial}{\partial z}\right)\hat{\Omega}_c = i \left( \dfrac{\Gamma \alpha}{2L}\right)  \hat{\sigma}_{23}, \label{oc}
\end{eqnarray}   
where $ L $ is the medium length, and $ \alpha = 4g^2 N L/c\Gamma$ is the optical density of atomic medium. $ N $ is the total atomic numbers in the ensemble.

Together with atomic equations in Eqs.~(\ref{S31}-\ref{S13}) and field equations in Eqs.~(\ref{op}, \ref{oc}), we have a set of coupled equations between atomic and field operators. To calculate entanglement properties between two fields, we apply the mean-field approximation, dividing each operator $ \hat{A} $ into two parts, i.e., $ \hat{A} = A + \hat{a} $, where $ A $ represents the mean-field value and $ \hat{a} $ corresponds to the quantum fluctuation operator.
Thus, we can decompose atomic and field operators as $ \hat{\sigma}_{\mu\nu} = \sigma_{\mu\nu} + \hat{s}_{\mu\nu},~\mu,\nu \in 1,2,3 $, and $ \hat{\mathcal{E}}_{\mu} = \mathcal{E}_\mu + \hat{a}_\mu,~\mu\in p,c $, where $ \hat{s}_{\mu\nu} $ and $ \hat{a}_{\mu} $ are dimensionless atomic and field fluctuation operators, respectively. 
The detail derivations are given in Appendix. 
 
In order to quantify the entanglement between two fields, we use Duan's 
inseparability \cite{criterion1, criterion2, criterion-EIT}, which is a sufficient condition for continuous-variable entanglement demonstrated by many experiments, i.e.,
\begin{eqnarray}
V(\theta) \equiv \Delta^2\left( \hat{X}_p + \hat{X}_c\right)(\theta) + \Delta^2\left( \hat{Y}_p - \hat{Y}_c\right)(\theta) < 4,~
\end{eqnarray}
where
$ \hat{X}_\sigma = \hat{a}_\sigma e^{-i\theta} + \hat{a}_{\sigma}^{\dagger} e^{i\theta}$ and $ \hat{Y}_\sigma = -i(\hat{a}_\sigma e^{-i\theta}- \hat{a}_{\sigma}^{\dagger}e^{i\theta} ) $ are the two quadrature operators of fields $ \hat{a}_{\sigma} $, $ \sigma\in {p, c} $, with the quadrature angle $ \theta$.
Expressing $ V(\theta)$ in terms of field operators, we have
\begin{eqnarray}
V(\theta) = 4\left[1+\langle\hat{a}_p^{\dagger}\hat{a}_p\rangle + \langle\hat{a}_c^{\dagger}\hat{a}_c\rangle + 2 \text{Re}\left( \langle\hat{a}_p\hat{a}_c\rangle e^{-2i\theta}\right) \right].\nonumber\\
\end{eqnarray}
By scanning all quadrature angles, one can find an optimum quadrature angle $ \theta_{\text{opt}} $, which minimizes the entanglement quantity $ V $. The entanglement quantity $ V(\theta) $ at $ \theta_{\text{opt}} $ is given by
\begin{eqnarray}
V = 4\left[ 1+\langle\hat{a}_p^{\dagger}\hat{a}_p\rangle + \langle\hat{a}_c^{\dagger}\hat{a}_c\rangle - 2 \mid\langle\hat{a}_p\hat{a}_c\rangle\mid\right], \label{Vopt}
\end{eqnarray}
while $ \theta_{\text{opt}} = \left( \text{Arg}\left[\langle \hat{a}_p\hat{a}_c\rangle\right]\pm n\pi\right) /2 $, and $ n \in $ odd.

The entanglement degree depends on some tunable parameters. 
In Sec. III, we will show the results of entanglement under various physical quantities, and compare the corresponding entanglement degree.

\section{iii. Numerical Results}

According to the theoretical model in Sec. II, we know that the entanglement is the function of optical density ($ \alpha $), two-photon detuning ($ \delta $), input Rabi-frequency of two fields ($ \Omega_{p,c} $), and the decoherence rate ($ \gamma_p $), which are measurable physical quantities in experiments. For simplicity, we consider asymmetric one-photon detuning, which is arranged as $ \Delta_p = -\Delta_c = \delta/2 $.  In this section, we will compare the two entanglement generation processes: one is by decoherence rate, and the other one is by two-photon detuning.
All the results in this section are obtained numerically. 

\FigTwo

In Fig.~\ref{fig2}(a), we have shown the relation between entanglement quantity $ V $ and the decoherence rate $ \gamma_p $. For given optical density $ \alpha $ and input Rabi frequencies of probe and coupling fields $ \Omega_{p,c} $, we can find an optimum decoherence rate $ \gamma_{p, \text{opt}} $ to maximize the output entanglement (the minimum value of $ V $, i.e., $ V_{\text{opt}, \gamma_p} $). 
If we give an optical density and a decoherence rate, there exists the optimum input Rabi frequency of coupling field $ \Omega_{c, \text{opt}} $, such that the output entanglement is maximum ($ V_{\text{opt}, \Omega_c} $), as shown in Fig.~\ref{fig2}(b). EIT condition in Fig.~\ref{fig2} has been used by setting $ \Omega_p = 0.1 \Omega_c $, and we consider on-resonance case, i.e., two-photon detuning $ \delta = 0 $.

Similarly, we consider how the two-photon detuning influences output entanglement. As shown in Fig.~\ref{fig3}(a), we can obtain the maximum entanglement by scanning two-photon detuning $ \delta $ for given optical density and input Rabi frequencies of probe and coupling fields. It is shown that one can find the optimum two-photon detuning $ \delta_{\text{opt}} $ and the corresponding entanglement quantity $ V_{\text{opt}, \delta} $. With the same process, there exists an optimum $ \Omega_{c,\text{opt}} $ to maximize output entanglement, which is $ V_{\text{opt}, \Omega_c} $ for given $ \alpha $ and $ \delta $, as shown in Fig.~\ref{fig3}(b). As Fig.~\ref{fig2}, we have set $ \Omega_p = 0.1 \Omega_c $ in order to satisfy EIT condition, and we let all the decoherence rate being zero to ensure that the entanglement is fully coming from the influence of $ \delta $. 

\FigThree

From Fig.~\ref{fig2}(a) and Fig.~\ref{fig3}(a), we can find that no entanglement is generated at output when $ \gamma_p  = 0 = \delta $, and entanglement between two fields is generated in the presence of $ \gamma_p $ or $ \delta $. 
Compared with the entanglement generated by the two processes, it is clear to see that the entanglement degree is larger and more efficient in two-photon detuning scheme.
In addition to the factors of $ \gamma_p $, $ \delta $, and $ \Omega_c $, entanglement also depends on the optical density, which is tunable and available in experiments. We are interested in maximum entanglement at different values of $ \alpha $.  
Figures ~\ref{fig4}(a) and 4(b) show the results based on the scheme by using $ \gamma_p $, and the results of $ \delta $ scheme are depicted in Fig.~\ref{fig4}(c,d) under various $ \alpha $'s. 
Figure ~\ref{fig4}(a) illustrates $ V_{\text{opt},\Omega_c} $ as a function of $ \gamma_p $, and the values of $ V_{\text{opt},\Omega_c} $ gradually glows with the increasing of $ \gamma_p $. For larger optical densities, it shows that the entanglement values degrades quickly. In Fig.~\ref{fig4}(b), it shows $ V_{\text{opt},\gamma_p} $, whose value gradually becomes larger with the increasing $ \Omega_c $, but
is insensitive to $ \alpha $. The entanglement values are changing from 3.8 to 4. 
In contrast, Figs.~\ref{fig4}(c) and 4(d) illustrate $ V_{\text{opt}, \Omega_c} $ and $ V_{\text{opt}, \delta} $ as the functions of $ \delta $ and $ \Omega_c $, respectively. The values of $ V $'s are also insensitive to the corresponding variables, but change significantly with $ \alpha $'s. The black dotted, green dashed-dotted, blue solid, and red dashed lines represent the values of $ \alpha $'s given by 100, 300, 1000, and 3000, respectively. From Fig.~\ref{fig4}(c,d), it manifests that the entanglement degree increases when optical density is increasing.
The optical density can enhance the output entanglement in the two-photon detuning scheme. 

\FigFour
Since the entanglement degree is much larger by using two-photon detuning scheme, we focus on the results of Fig.~\ref{fig4}(c,d). 
We can see that the value of optimum entanglement, $ V_{\text{opt},\Omega_c} $ and $ V_{\text{opt},\delta} $, under a given optical density $ \alpha $ is almost a constant.

\FigFive
In Fig.~\ref{fig5}, we have numerically plotted the contour plot of entanglement quantity $ V $ with respect to $ \Omega_p $ and $ \delta $ under three different $ \Omega_c $'s, which are $ 0.85\Gamma $, $ 1.2\Gamma $, and $ 1.7\Gamma $, respectively. As shown in the three plots, we can see that the values of $ V $ are the same, but with different ranges of $ \Omega_p $ and $ \delta $. The positions of $ \Omega_{p} $ corresponding to the same values of $ V $ is clearly proportional to $ \Omega_c $. Similarly, the positions of $ \delta $ is proportional to $ \Omega_c^2 $.
It implies that there exists a relationship among entanglement quantity $ V $ and the ratios of $ \Omega_p/\Omega_c $ and $ \delta/\Omega_c^2 $. The deeper understanding to the results in Fig.~\ref{fig5} will be discussed in Sec. IV. 

%%%%%%%%%%%%%%%%%%%%%%%%%%%%%%%%%%%%%%%%%%%%%%%%%%%%%%%%%%%%%%%%%%%%%%%%%%
\section{iv. Discussions}

Generation of CV quantum entanglement between probe and coupling fields using atomic EIT system is the key point of this paper. We have proposed a theoretical model in Sec. II, deriving equations of motion for atomic and field operators, and showing the numerical simulation results in Sec.III. In order to understand the physics behind these results thoroughly, in this section we study the system analytically from the framework given in Sec. II. Using Eqs.~(\ref{S31} - \ref{oc}), one can obtain 
\begin{eqnarray}
&\dfrac{\partial}{\partial \zeta}\hat{a}_p = P_1 \hat{a}_p + Q_1 \hat{a}_p^{\dagger} + R_1 \hat{a}_c + S_1 \hat{a}_c^{\dagger} + \hat{n}_p, \\
&\dfrac{\partial}{\partial \zeta}\hat{a}_c = P_2 \hat{a}_p + Q_2 \hat{a}_p^{\dagger} + R_2 \hat{a}_c + S_2 \hat{a}_c^{\dagger} + \hat{n}_c,
\label{EQ}
\end{eqnarray}
where $ \zeta \equiv z/L $ is the dimensionless length, and $ \hat{n}_p $ and $ \hat{n}_c $ are the corresponding Langevin noise operators.     
$ P_i $, $ Q_i $, $ R_i $, and $ S_i $ ($ i = 1,2 $) are the coefficients.
For two-photon detuning scheme, i.e., $ \gamma_p = 0 $, we have
\begin{eqnarray}
\begin{split}
&P_1 \simeq iK - \lambda ~,~Q_1 \simeq -2i\mu re^{2iK\zeta}, \\
&R_1\simeq -i\mu e^{iK\zeta}~,~S_1 \simeq-i\mu e^{iK\zeta},\\
&P_2 \simeq -i\mu e^{-iK\zeta}~,~Q_2\simeq -i\mu e^{iK\zeta},\\
&R_2 \simeq i\mu r~,~S_2\simeq2i\mu r,
\end{split}
\label{coe}
\end{eqnarray} 
where $ K\equiv \alpha\epsilon $ and $ \lambda \equiv 2\alpha\epsilon^2 $, which are the imaginary and real part of $ P_1 $. 
$ \mu\equiv\alpha\epsilon r $, and
$ r \equiv \vert\Omega_p/\Omega_c \vert $ which is the ratio between probe and coupling Rabi frequencies, being much smaller than 1 under EIT condition i.e., $ r \ll 1 $. 
We also have  $\epsilon\equiv\Gamma\delta/\Omega_c^2 $.
In our analytical study, we assume that the amplitudes of probe and coupling fields are unchanged. It means that $ r $ is a constant. 
Moreover, we don not consider the phase of coupling field because the phase change is very small. Thus, $ \Omega_c $ is real in our case. 
On the contrary, we have to take the probe field phase into account.
The phase of probe field is $ K\zeta $, which can be caught from the coefficients shown in Eq. (\ref{coe}).
The phase term can be eliminated by transforming field operators into a new rotating frame by defining $ \hat{O} \rightarrow \hat{O}e^{iK\zeta} $, where $ \hat{O} $ represents $ \hat{a}_p, \hat{a}_p^{\dagger}, \hat{n}_p, \text{and} ~ \hat{n}_p^{\dagger} $.

Now we turn to consider the entanglement between probe and coupling fields. 
From Eq. (\ref{coe}), it is clearly to see that the terms of $ R_1$ and $S_1 $ link coupling field operators ($ \hat{a}_c $, $\hat{a}_c^{\dagger}$) and probe field operators ($ \hat{a}_p $, $\hat{a}_p^{\dagger}$); while $ P_2$ and $Q_2 $ make the correlation between probe field operators and coupling field operators.
On the other hand, the coefficients of $ P_1,~Q_1,~ R_2,~\text{and}~S_2 $ correspond to the self-interaction processes for each field. 
Let's discuss the physical meaning of these coefficients. First, the probe-coupling entanglement is mainly produced by the coefficients $S_1$ and $Q_1$. If there only exists $S_1$ and $Q_2$, due to their phase terms, $e^{iK\zeta}$, the output entanglement $V$ is oscillating
between 4 and 2.
Second, the physics of the terms of $ R_1 $ and $ P_2 $ is the cross-phase coupling, which can't produce entanglement at output. Third, when we only consider the coefficients of $ Q_1 $ and $ S_2 $, which correspond to the single-mode squeezing processes, we can obtain the two independent squeezed lights. Finally, the coefficients of $ P_1 $ and $ R_2 $ are related to self-phase or damping/amplification processes, which also do not have the abilities to produce entanglement. 

In order to obtain the analytical expression for entanglement, we consider the coefficients coming from the $ 0^{\text{th}} $ and $ 1^{\text{st}} $ order terms of $ r $, i.e., $ P_1,~R_1,~S_1, P_2 $, and $ Q_2 $, as well as the Langvin noise contributions from $ \hat{n}_p $ and $ \hat{n}_c $. We yield a form as follows.
\begin{eqnarray}
V_1 = 4\left[ 1 + \mu^2\left( \dfrac{e^{-2\lambda} + 2\lambda - 1}{\lambda^2}\right)  - 2\mu\left( \dfrac{1-e^{-\lambda}}{\lambda}\right) \right], \label{V_ana1}
\end{eqnarray}

When $ \lambda \rightarrow 0 $, the main contributions are coming from $ R_1, S_1, P_2 $, and $ Q_2 $, resulting in the entanglement given by
$ V_1 = 4\left(1+2\mu^2 -2\mu \right)  $, which only depends on $ \mu $.
It implies that $ V_{\text{best}} = 2 $ when $ \mu = 1/2 $, which means the entanglement degree is independent of $ r $ as long as the condition $ \mu = 1/2 $. It is quite different from the case of the entanglement by two-mode squeezing, which can approach to an ideal entangled state as $ \mu\rightarrow \infty $.

According to Fig.~\ref{fig5}, we can plot the entanglement quantity $ V $ with respect to $ r $ and $ \epsilon $. 
After plotting with the new variables, we can find that the three plots of Fig.~\ref{fig5}(a - c) correspond to the same result shown in Fig.~\ref{fig6}.  
It implies that $ V $ depends only on two independent parameters of $ \epsilon $ and $ r $, i.e., as long as $ \epsilon $ and $ r $ are given. Any combination of $ \delta $, $ \Omega_c $, and $ \Omega_p $ results in the same value of $ V $. One can clearly see that in the plot the condition of $ \mu = 1/2 $, represented by the dashed line of a hyperbolic function, crosses the minimum or optimum value of $ V $.

\FigSix

However, $ \mu = 1/2 $ is not a sufficient condition to find the optimum entanglement, and Eq.~(\ref{V_ana1}) can't explain our results completely. The main reason for this problem is that the higher order terms of $ r $ become important when $ r $ is getting large. 
From Eq.~(\ref{coe}), one can see that $ R_2 $, $ Q_1 $ and $ S_2 $ are proportional to $ r^2 $. 
However, the coefficient $ R_2 $ is actually the phase of coupling field, which can be eliminated by using the transformation of $ \hat{a}_c \rightarrow \hat{a}_c e^{+i\mu r\zeta} $. 
Thus, we only consider the influence of $ r^2 $ from the terms of $ Q_1 $ and $ S_2 $, which are related to the single-mode squeezing coefficient.
According to Eq.~(\ref{Vopt}), one can see that the entanglement degree depends on the average photon numbers of probe and coupling fields.
For single-mode squeezed state of probe and coupling, the average photon numbers are $ \sinh^2(\vert Q_1\vert\zeta) $ and $ \sinh^2(\vert S_2\vert\zeta) $ rather than 0. As a reason, we can naively modify output entanglement by considering the external photon numbers coming from the single-mode squeezing terms, but without introducing the corresponding extra noises. Thus it reads as
\begin{eqnarray}
\label{V_ana2}
&&V \simeq V_1 + 8\sinh^2\left( 2\mu r\right).
\end{eqnarray}
 
For the case $ \lambda  \ll 1 $, we can expand $ V_1 $ to $\mathcal{O(\lambda)} $. With these approximations, we can obtain a closed form of output entanglement shown below: 
\begin{eqnarray}
\label{V_close_form}
V \simeq 4\left( 1 + 2\mu^2 - 2\mu\right) + 4\mu\lambda\left( 1-4\mu/3\right) + 8 (2\mu r)^2 .
\end{eqnarray}
From Eq.~(\ref{V_close_form}), we can obtain the optimum entanglement by substituting $ \mu = 1/2 $, and express $ V $ as the function of $ \alpha $, $ \epsilon $, and $ r $ by using $ \lambda = 2\alpha\epsilon^2 $. It will be
\begin{eqnarray}
V_{\text{opt}} = 2 + \dfrac{4}{3}\alpha\epsilon^2 + 8r^2.
\label{V_Lr}
\end{eqnarray}

Eq. (\ref{V_Lr}) is the optimum entanglement by considering coefficients of
$ P_1 $, $ R_1 $, $ S_1 $, $ P_2 $, $ Q_2 $, $ Q_1 $, and $ S_2 $. 
From Eq. (\ref{V_Lr}), we can find the best entanglement by 
using Lagrangian multiply with the constraint condition given by $ \mu = 1/2 $. It shows that 
\begin{eqnarray}
&&\epsilon_{\text{opt}} = (3/2)^{1/4}\alpha^{-3/4},\label{epsilon_opt}\\
&&r_{\text{best}} = (24\alpha)^{-1/4}.
\label{r_best}
\end{eqnarray}
From Eqs.~(\ref{epsilon_opt}, \ref{r_best}), we can see that $ \epsilon_{\text{opt}} $ and $ r_{\text{best}} $ are constants when $ \alpha $ is given. Under these conditions, the best entanglement value is
\begin{eqnarray}
V_{\text{best}} = 2 + (32/3)^{-1/2}\alpha^{-1/2}, 
\label{Vbest}
\end{eqnarray} 
which only depends on optical density $ \alpha $. 
The result reflects the fact that the value of $ \text{log}_{10}(V-2) $ would decrease $ 0.5 $ with the increment of an order of magnitude in optical density. It quantitatively matches the results shown in Fig.~\ref{fig4}(c, d).     

In comparison with the entanglement generation from decoherence rate, we have seen that the scheme of two-photon detuning is more efficient from Fig.~\ref{fig4}. The physics behind the result can be understood as follows.
Because the entanglement degree is mainly contributed from the coefficients $ S_1 $ and $ Q_2 $, we study how the two coefficients affect the entanglement quantity. 
For the scheme of the decoherence rate, the entanglement coefficients are $ \vert S_1\vert= \vert Q_2\vert = \alpha\varepsilon r e^{-\alpha\varepsilon\zeta} $, where $ \varepsilon\equiv\Gamma\gamma_p/\vert\Omega_c\vert^2 $. 
Unlike the detuning scheme, the decay factors, $e^{-\alpha\varepsilon\zeta}$, in $S_1$ and $Q_2$ introduce noise into the system, and the degree of entanglement can only go as far as a little below 4.
On the other hand, in detuning case, $S_1$ and $Q_2$ have the phase term, $e^{i\alpha\epsilon\zeta}$ (where $\epsilon = \Gamma\delta/|\Omega_c|^2$), instead of the decay term. A larger value of $\alpha\epsilon r$ neither causes any decay nor introduces more noise into the system.
In the scenario, the contribution from $ S_1 $ and $ Q_2 $ is proportional to $ \alpha\epsilon r $.
The result implies that the entanglement degree can be enhanced by optical density in two-photon detuning scheme. In contrast, the entanglement degree in decoherence rate scheme becomes worse for larger optical density, as shown in Fig.~\ref{fig4}(a,~b).  

The existence of the optimum ratio $ r $ is coming from the competition between dissipation and single-mode squeezing. From Eq.~(\ref{V_Lr}), using the constraint $ \mu = 1/2 $, we can rewrite that the optimum entanglement is the function of $ r $ as shown as follows. 
\begin{eqnarray}
V_{\text{opt}} = 2 + (3\alpha)^{-1}r^{-2} + 8r^2. \label{V_opr_r}
\end{eqnarray}
From Eq.~(\ref{V_opr_r}), it is clearly to see that 
the $ r $-dependent entanglement is the result of the sum of the two terms shown in the second and third term.  
The second term is coming from $ \lambda $, which is related to the dissipation term of probe field. It will introduce extra noise into the system.
On the other hand, the third terms are attributed to single-mode squeezing term.
When $ r $ is small, the extra noise dominates the output entanglement value, while the effect of single-mode squeezing term becomes important when $ r $ is getting larger. As a result, there exists an optimum value of $ r $ to minimize entanglement value $ V $. 
On the other hand, the optimum entanglement can be expressed as the function of $ \epsilon $ as given by
 $ V_{\text{opt}} = 2 + (4\alpha/3)\epsilon^2 + (2/\alpha^2)\epsilon^{-2} $. It implies that there exists a best entanglement when the sum of extra noise from dissipation and single-mode squeezing is minimized. 
Generally, two-mode squeezing ($ S_1,~Q_2 $) and the cross-coupling terms ($ R_1,~P_2 $) and the imaginary part of $ P_1 $ limit the best entanglement to be 2, which is $ 50\% $ of ideal entangled state. 
The real part of $ P_1 $ associating with the dissipation term will introduce extra noise fluctuation which is proportional to $ r^{-2} $, and degrade the output entanglement in the region of $ r\ll 1 $.
In contrast, the single-mode squeezing, $ Q_1 $ and $ S_2 $, will degrade the entanglement degree and is proportional to $ r^2 $, which destroys entanglement when $ r $ is getting larger. 
Similarly, we also have optimum $ \epsilon $ from the condition of $ \mu = 1/2 $. As a result, we can find the best strength ratio $ r $ and detuning-coupling field ratio $ \epsilon $ to minimize the output entanglement.

\section{v. Conclusion}

In the present work, we have discussed the generation of quantum entanglement between probe and coupling fields under EIT condition.
We compare the entanglement degree arising from two different mechanisms, which are decoherence rate and two-photon detuning.
Our study has identified that it is more efficient to obtain higher output entanglement degree by introducing two-photon detuning.  
Furthermore, we have numerically studied the influence of the EIT parameters, which are two-photon detuning, field Rabi frequencies, and optical density to entanglement degree. Also, the conditions of the corresponding parameters for obtaining the optimal entanglement have been found from theoretical analysis. 
It shows that the two-mode squeezing and cross-coupling terms give us a constraint for the parameters to obtain the best entanglement, i.e. $ \mu = 1/2 $. 
The noise fluctuation  from probe field dissipation and the single-mode squeezing from probe and coupling fields will reduce the entanglement degree.
The optimum condition of $ r $ and $ \epsilon $ for the best entanglement have been found.
The study contributes to our understanding of  the origin of entanglement induced by atom-field interaction in EIT system, as well as a deeper connection between quantum coherence and entanglement. The work can be further extended to more complicated atomic systems, which have possibilities to produce higher entanglement degree, conducing the progresses in the development of CV quantum information sciences. 

\section{acknowledgement}

This work was supported by the Ministry of Science and Technology of Taiwan under Grant Nos. 105-2628-M-007-003-MY4, 106-2119-M-007-003, and 107-2745-M-007-001.
\begin{widetext}

%%%%%%%%%%%%%%%%%%%%%%%%%%%%%%%%%%%%%%%%%%%%%%%%%%%%%%%%%%%%%%%%%%%%%%%
\renewcommand{\thesection}{\mbox{Appendix~\Roman{section}}} %
\setcounter{section}{0}
\renewcommand{\theequation}{\mbox{A.\arabic{equation}}} %\section{Appendix}
\setcounter{equation}{0} 
\appendix*
\onecolumngrid
\section{Appendix}

In this Appendix, we will derive the equations of motion for quantum fluctuations of atomic operators given by Eqs.~(\ref{S31}-\ref{S13}) as well as the field fluctuations given in Eq.~(\ref{op}, \ref{oc}). 
By using the mean-field approximation, we can decompose an operator into two parts, which are mean-field part and the corresponding quantum fluctuation part, and one can obtain linear equations for fluctuation operators by ignoring the higher-order fluctuation terms.   
In the following, we have shown the linearized eqautions of atomic fluctuation operators. 

\begin{eqnarray}
&&\dfrac{\partial}{\partial t}\hat{s}_{31} = -\left( \dfrac{\Gamma}{2}+i\Delta_p\right)  \hat{s}_{31}-\dfrac{i}{2}\Omega_p^{\ast}(\hat{s}_{11}-\hat{s}_{33})-\dfrac{i}{2}g(\sigma_{11}-\sigma_{33})\hat{a}_p^{\dagger} 
- \dfrac{i}{2}\Omega_c^{\ast}\hat{s}_{21}-\dfrac{i}{2}g\sigma_{21}\hat{a}_c^{\dagger}+\hat{F}_{31}\label{s31},\\
&&\dfrac{\partial}{\partial t}\hat{s}_{32} = -\left( \dfrac{\Gamma}{2}+i\Delta_c\right) \hat{s}_{32}-\dfrac{i}{2}\hat{\Omega}_c^{\ast}(\hat{s}_{22}-\hat{s}_{33})-\dfrac{i}{2}g(\sigma_{22}-\sigma_{33})\hat{a}_c^{\dagger}- \dfrac{i}{2}\Omega_p^{\ast}\hat{s}_{12}-\dfrac{i}{2}g\sigma_{12}\hat{a}_p^{\dagger}
+\hat{F}_{32}\label{s32},\\
&&\dfrac{\partial}{\partial t}\hat{s}_{21} = -\left( {\gamma}_p+i\delta\right) \hat{s}_{21}+\dfrac{i}{2}\Omega_p^{\ast}\hat{s}_{23}+\dfrac{i}{2}g\sigma_{23}\hat{a}_p^{\dagger}- \dfrac{i}{2}\sigma_{31}\hat{u}_c - 
\dfrac{i}{2}\Omega_c\hat{s}_{31}+\hat{F}_{21}\label{s21},\\
&&\dfrac{\partial}{\partial t}\hat{s}_{11} = \Gamma_1 \hat{s}_{33}-\dfrac{i}{2}\Omega_p\hat{s}_{31}-\dfrac{i}{2}g\sigma_{31}\hat{a}_p+ \dfrac{i}{2}\Omega_p^{\ast}\hat{s}_{13}+\dfrac{i}{2}g\sigma_{13}\hat{a}_p^{\dagger}+\hat{F}_{11}\label{s11},\\
&&\dfrac{\partial}{\partial t}\hat{s}_{22} = \Gamma_2 \hat{s}_{33}-\dfrac{i}{2}\Omega_c \hat{s}_{32}-\dfrac{i}{2}g\sigma_{32}\hat{a}_c+\dfrac{i}{2}\Omega_c^{\ast}\hat{s}_{23}+\dfrac{i}{2}g\sigma_{23}\hat{a}_c^{\dagger}+\hat{F}_{22}\label{s22},\\
&&\dfrac{\partial}{\partial t}\hat{s}_{33} = -\Gamma \hat{s}_{33}
+\dfrac{i}{2}\Omega_p\hat{s}_{31}+\dfrac{i}{2}g\sigma_{31}\hat{a}_p+\dfrac{i}{2}\Omega_c\hat{s}_{32}+\dfrac{i}{2}g\sigma_{32}\hat{a}_c -\dfrac{i}{2}\Omega_p^{\ast}\hat{s}_{13}
-\dfrac{i}{2}g\sigma_{13}\hat{a}_p^{\dagger}
-\dfrac{i}{2}\Omega_c^{\ast}\hat{s}_{23}
-\dfrac{i}{2}g\sigma_{23}\hat{a}_c^{\dagger}+\hat{F}_{33}, \label{s33}\nonumber\\
&& \\
&&\dfrac{\partial}{\partial t}\hat{s}_{12} = -\left( {\gamma}_p-i\delta\right) \hat{s}_{12}-\dfrac{i}{2}\Omega_p\hat{s}_{32}
-\dfrac{i}{2}g\sigma_{32}\hat{a}_p+ \dfrac{i}{2}\sigma_{13}\hat{u}_c^{\dagger} + \dfrac{i}{2}\Omega_c^{\ast}\hat{s}_{13}+\hat{F}_{12},\label{s12}\\
&&\dfrac{\partial}{\partial t}\hat{s}_{23} = -\left( \dfrac{\Gamma}{2}-i\Delta_c\right) \hat{s}_{23}+\dfrac{i}{2}\hat{\Omega}_c(\hat{s}_{22}-\hat{s}_{33})+\dfrac{i}{2}g(\sigma_{22}-\sigma_{33})\hat{a}_c
+ \dfrac{i}{2}\Omega_p\hat{s}_{21}+\dfrac{i}{2}g\sigma_{21}\hat{a}_p+\hat{F}_{23},\label{s23}\\
&&\dfrac{\partial}{\partial t}\hat{s}_{13} = -\left( \dfrac{\Gamma}{2}-i\Delta_p\right)  \hat{s}_{13}+\dfrac{i}{2}\Omega_p(\hat{s}_{11}-\hat{s}_{33})+\dfrac{i}{2}g(\sigma_{11}-\sigma_{33})\hat{a}_p 
+ \dfrac{i}{2}\Omega_c\hat{s}_{12}+\dfrac{i}{2}g\sigma_{12}\hat{a}_c+\hat{F}_{13},\label{s13}
\end{eqnarray}  

where $ \hat{s}_{\mu\nu} = \hat{s}_{\nu\mu}^{\dagger}$, $ \mu, \nu\in 1,2,3 $. Essentially, Eqs.~(\ref{s12} - \ref{s13}) are the hermitian conjugate equations of Eq.~(\ref{s31} - \ref{s21}).
Since we are interested in the steady-state solution for output field operators, we can set the time derivative to be zero.
Thus, we can express Eqs.~(\ref{s31} - \ref{s13}) in the matrix form as 
$\textbf{M}_\textbf{1}~\textbf{y} + \textbf{M}_\textbf{2}~\textbf{u} + \textbf{r} = 0$, where $ \textbf{y}^{T} = \left( \hat{s}_{31},  \hat{s}_{32},  \hat{s}_{21},  \hat{s}_{11},  \hat{s}_{22}, \hat{s}_{33}, \hat{s}_{12}, \hat{s}_{23}, \hat{s}_{13} \right)$ gives  the fluctuations of atomic operators,  $\textbf{a}^{T} = \left( \hat{a}_p,  \hat{a}_p^{\dagger}, \hat{a}_c,  \hat{a}_c^{\dagger} \right)$ denotes the fluctuations of field operators, and $\textbf{r}^{T} = \left( \hat{F}_{31},  \hat{F}_{32},  \hat{F}_{21},  \hat{F}_{11},  \hat{F}_{22}, \hat{F}_{33}, \hat{F}_{12}, \hat{F}_{23}, \hat{F}_{13} \right)$ is the corresponding Langevin noise operators, respectively. The matrices $ \textbf{M}_\textbf{1} $ and $ \textbf{M}_\textbf{2} $ are 9 by 9 and 9 by 4 matrix. 
We have shown the matrix expression as follows.

\begin{eqnarray}
\textbf{M}_\textbf{1} = \left( \begin{array}{ccccccccc} 
-\left( \dfrac{\Gamma}{2}+i\Delta_p\right)   & 0 & -i\dfrac{\Omega_c^{\ast}}{2} & -i\dfrac{\Omega_p^{\ast}}{2} & 0 & i\dfrac{\Omega_p^{\ast}}{2} & 0 & 0 & 0 \\
0 & -\left( \dfrac{\Gamma}{2}+i\Delta_c\right)  & 0 & 0 & -i\dfrac{\Omega_c^{\ast}}{2} & i\dfrac{\Omega_c^{\ast}}{2} & -i\dfrac{\Omega_p^{\ast}}{2} & 0 & 0\\
-i\dfrac{\Omega_c}{2} & 0 & -\left( \gamma_p+i\delta\right)  & 0 & 0 & 0 & 0 & i\dfrac{\Omega_p^{\ast}}{2} & 0\\
-i\dfrac{\Omega_p}{2} & 0 & 0 & 0 & 0 & \dfrac{\Gamma}{2} & 0 & 0 & i\dfrac{\Omega_p^{\ast}}{2} \\
0 & -i\dfrac{\Omega_c}{2} & 0 & 0 & 0 & \dfrac{\Gamma}{2} & 0 & i\dfrac{\Omega_c^{\ast}}{2} & 0 \\
0 & 0 & 0 & 1 & 1 & 1 & 0 & 0 & 0 \\
0 & -i\dfrac{\Omega_p}{2} & 0 & 0 & 0 & 0 & -\left( \gamma_p-i\delta\right)  & 0 & i\dfrac{\Omega_c^{\ast}}{2}\\
0 & 0 & i\dfrac{\Omega_p}{2} & 0 & i\dfrac{\Omega_c}{2} & -i\dfrac{\Omega_c}{2} & 0 & -\left( \dfrac{\Gamma}{2}-i\Delta_c\right)  & 0\\
0 & 0 & 0 & i\dfrac{\Omega_p}{2} & 0 & -i\dfrac{\Omega_p}{2} & i\dfrac{\Omega_c}{2} & 0 & -\left( \dfrac{\Gamma}{2}-i\Delta_p\right) 
\end{array}\right)_{9\times 9} \nonumber
\end{eqnarray}
\begin{eqnarray}
\textbf{M}_\textbf{2} = \dfrac{g}{2}\left( \begin{array}{cccc} 
0 & -i\left( \sigma_{11}-\sigma_{33}\right) & 0 & -i\sigma_{21} \\
0 & -i\sigma_{12} & 0 & -i\left( \sigma_{22}-\sigma_{33}\right) \\
0 & i\sigma_{23} & -i\sigma_{31} & 0 \\
-i\sigma_{31} & i\sigma_{13} & 0 & 0 \\
0 & 0 & -i\sigma_{32} & i\sigma_{23} \\
0 & 0 & 0 & 0 \\
-i\sigma_{32} & 0 & 0 & i\sigma_{13} \\
i\sigma_{21} & 0 & i\left( \sigma_{22}-\sigma_{33}\right) & 0 \\
i\left( \sigma_{11}-\sigma_{33}\right) & 0 & i\sigma_{12} & 0
\end{array}\right)_{9\times 4}. \nonumber
\end{eqnarray}
where $ \Gamma_1 = \Gamma_2 = \Gamma/2 $ has been used.
The atomic fluctuation operators can be easily expressed in terms of filed operators by solving
$ \textbf{y} = \textbf{T}\left( \textbf{M}_2 \textbf{a} + \textbf{r}\right) $, where $ \textbf{T} \equiv -\textbf{M}_1^{\text{T}} $.

On the other hand, the field fluctuation equations under steady-state regime are given by 
\begin{eqnarray}
&&\dfrac{\partial}{\partial \zeta} \hat{a}_p = i \left( \dfrac{\Gamma \alpha}{2g}\right)  \hat{s}_{13}, \label{app_op}\\  
&&\dfrac{\partial}{\partial \zeta} \hat{a}_c = i \left( \dfrac{\Gamma \alpha}{2g}\right)\hat{s}_{23}. \label{app_oc}
\end{eqnarray} 
where $ \zeta\equiv z/L $, which is the normalized distance.
The source terms on right-hand side coming from the atomic coherence operators, which can be directly replaced by field fluctuation operators, i.e. 
$ \hat{s}_{13} = \hat{s}_{13}\left( \hat{a}_p, \hat{a}_p^{\dagger},\hat{a}_c, \hat{a}_c^{\dagger}\right) $, and $ \hat{s}_{23} = \hat{s}_{23}\left( \hat{a}_p, \hat{a}_p^{\dagger},\hat{a}_c, \hat{a}_c^{\dagger}\right) $. 
The general expressions of $ \hat{s}_{13} $ and $ \hat{s}_{23} $ are given as
\begin{eqnarray}
\hat{s}_{13} = A_1 \hat{u}_p + B_1 \hat{u}_p^{\dagger} + C_1 \hat{u}_c + D_1 \hat{u}_c^{\dagger} + \hat{f}_{13}\label{app_s13},\\
\hat{s}_{23} = A_2 \hat{u}_p + B_2 \hat{u}_p^{\dagger} + C_2 \hat{u}_c + D_2 \hat{u}_c^{\dagger} + \hat{f}_{23}\label{app_s23}.
\end{eqnarray}
where $ \hat{f}_{13} $ and $ \hat{f}_{23} $ are the effective Langevin noise operator from Langevin noise operators $ \hat{F}_{\mu\nu} $'s given in Eq.(\ref{s31}-\ref{s13}).\\

Substituting Eq.~(\ref{app_s13}, \ref{app_s23} ) into Eqs.~(\ref{app_op}, \ref{app_oc}), we can obtain the field propagation equations for field fluctuation operators. The compact form is given as follows.
\begin{eqnarray}
\dfrac{\partial}{\partial\zeta}\textbf{a} = \textbf{C}\textbf{a} + \textbf{N}  \label{app_eqn_a}
\end{eqnarray}  
In which $\textbf{a}^{\textbf{T}} \equiv \left( \hat{a}_p, \hat{a}_p^{\dagger}, \hat{a}_c, \hat{a}_c^{\dagger}\right)$, and 
the two matrices of $\textbf{C} $ and $\textbf{N} $ have the explicit form as
\begin{eqnarray}
&&\textbf{C} = i\dfrac{\Gamma \alpha}{2}\left( \begin{array}{cccc}
A_1 & B_1 & C_1 & D_1 \\
-B_1^{\ast} & -A_1^{\ast} & -D_1^{\ast} & -C_1^{\ast} \\
A_2 & B_2 & C_2 & D_2 \\
-B_2^{\ast} & -A_2^{\ast} & -D_2^{\ast} & -C_2^{\ast} 
\end{array}\right)\equiv
\left( \begin{array}{cccc}
P_1 & Q_1 & R_1 & S_1 \\
Q_1^{\ast} & P_1^{\ast} & S_1^{\ast} & R_1^{\ast} \\
P_2 & Q_2 & R_2 & S_2 \\
Q_2^{\ast} & P_2^{\ast} & S_2^{\ast} & R_2^{\ast} 
\end{array}\right),\label{C}\\
&&\textbf{N} = i\dfrac{\Gamma \alpha}{2g}\left( \begin{array}{c}
\hat{f}_{13},
-\hat{f}_{13}^{\dagger},
\hat{f}_{23},
-\hat{f}_{23}^{\dagger}
\end{array}\right)^{T}.
\label{N}
\end{eqnarray}

The correlations between two field fluctuation operators can be calculated from Eq.~(\ref{app_eqn_a}). It is straightforwardly to have the form as follows.
\begin{eqnarray}
\dfrac{\partial}{\partial\xi}\textbf{S} = \textbf{C}\textbf{S} + \textbf{S}\textbf{C}^{\dagger} + \textbf{Z}.
\label{aad}
\end{eqnarray}
Here, $ \textbf{S}\equiv \langle\textbf{a}\textbf{a}^{\dagger}\rangle $, and the matrix $\textbf{Z}$ shows the correlations of Langevin noise operators, denoted  $\langle \textbf{N}\textbf{N}^{\dagger}\rangle$. That is 
\begin{eqnarray}
\textbf{Z}\equiv\langle \textbf{N}\textbf{N}^{\dagger}\rangle = \dfrac{\Gamma \alpha}{4} \left( \textbf{V} \mathcal{D} \textbf{V}^{\dagger}\right).
\label{NN}
\end{eqnarray}
Here, we have to consider the correlations of any two Langevin noise operators, i.e., $\langle\hat{F}_{\mu}\hat{F}_{\nu}\rangle = \mathcal{D}_{\mu\nu} \, c/(NL)$, in which the diffusion coefficient,  $\mathcal{D}_{\mu\nu}$ is given by
\begin{eqnarray}
\mathcal{D} = \left( \begin{array}{ccccccccc}
0 & 0 & \gamma_p\sigma_{32} & 0 & 0 & 0 & 0 & 0 & 0 \\
0 & 0 & 0 & 0 & 0 & 0 & \gamma_p\sigma_{31} & 0 & 0 \\
\gamma_p\sigma_{23} & 0 & 2\gamma_p\sigma_{22}+\Gamma_2\sigma_{33} & 0 & 0 & 0 & 0 & 0 & 0 \\
0 & 0 & 0 & \Gamma_1\sigma_{33} & 0 & 0 & 0 & -\Gamma_1\sigma_{32} & -\Gamma_1\sigma_{31} \\
0 & 0 & 0 & 0 & \Gamma_2\sigma_{33} & 0 & 0 & -\Gamma_2\sigma_{32} & -\Gamma_2\sigma_{31} \\
0 & 0 & 0 & 0 & 0 & 0 & 0 & 0 & 0 \\
0 & \gamma_p\sigma_{13} & 0 & 0 & 0 & 0 & 2\gamma_p\sigma_{11}+\Gamma_1\sigma_{33} & 0 & 0 \\
0 & 0 & 0 & -\Gamma_1\sigma_{23} & -\Gamma_2\sigma_{23} & 0 & 0 & \Gamma_2\sigma_{33}+\Gamma\sigma_{22} & (\Gamma-\gamma_p)\sigma_{21} \\
0 & 0 & 0 & -\Gamma_1\sigma_{13} & -\Gamma_2\sigma_{13} & 0 & 0 & (\Gamma-\gamma_p)\sigma_{12} & \Gamma_1\sigma_{33}+\Gamma\sigma_{11} 
\end{array}
\right)_{9\times 9} .
\label{rrd}
\end{eqnarray}
The matrix \textbf{V} is related to matrix $ \textbf{T} $, and has the form:
\begin{eqnarray}
\textbf{V} \equiv \left( \begin{array}{ccccccccc}
T_{91} & T_{92} & T_{93} & T_{94} & T_{95} & T_{96} & T_{97} & T_{98} &  T_{99} \\
-T_{11} & -T_{12} & -T_{13} & -T_{14} & -T_{15} & -T_{16} & -T_{17} & -T_{18} & -T_{19} \\
T_{81} & T_{82} & T_{83} & T_{84} & T_{85} & T_{86} & T_{87} & T_{88} &  T_{89} \\
-T_{21} & -T_{22} & -T_{23} & -T_{24} & -T_{25} & -T_{26} & -T_{27} & -T_{28} &  -T_{29} 
\end{array}\right)_{4\times 9},
\label{V}
\end{eqnarray}
~\\
By solving Eq.~(\ref{aad}), one can calculate the entanglement degree based on Eq.~(\ref{Vopt}) with the matrix elements of $ \textbf{S} $:
\begin{eqnarray}
V = 4\left( 1 + \textbf{S}_{22} + \textbf{S}_{44} -2\left\vert \textbf{S}_{14}\right\vert \right). 
\end{eqnarray}
~\\

\end{widetext}
%%%%%%%%%%%%%%%%%%%%%%%%%%%%%%%%%%%%%%%%%%%%%%%

\end{document}